%% file: vbf_h3j.tex
\documentclass[hyper,11pt,a4paper]{JHEP3}
\usepackage{epsfig}

\newlength{\largfig}
\def\beq{\begin{equation}} 
\def\eeq{\end{equation}} 
 
\def\beqn{\begin{eqnarray}} 
\def\eeqn{\end{eqnarray}} 

\def\slashit#1{\slash\mkern-10mu#1}
\def\Re{{\rm Re}}

\def\fortran{{\tt fortran}}

\def\MADGRAPH{{\tt MadGraph}}

\def\td#1{\widetilde{D}_{#1}}
\def\epsaepsb{\epsilon_1  \cdot \epsilon_2}
\def\epsakb{\epsilon_1 \cdot k_2}
\def\epsaqb{\epsilon_1 \cdot q_2}
\def\epsaqa{\epsilon_1 \cdot q_1}
\def\epsbkb{\epsilon_2 \cdot k_2}
\def\epsbqb{\epsilon_2 \cdot q_2}
\def\epsbqa{\epsilon_2 \cdot q_1}
\def\epsbka{\epsilon_2 \cdot  k_1}
\def\epsaka{\epsilon_1 \cdot  k_1}
\def\qbs{q_{2}^{2}}
\def\eps{\epsilon}
\def\mc{\mathcal}

\newcommand{\p}[1]{\ensuremath{p_#1}}
\newcommand{\pa}{\ensuremath{p_{a}}}
\newcommand{\pb}{\ensuremath{p_{b}}}

\newcommand{\pt}[1]{\ensuremath{\tilde{p}_{#1}}} 
 
\newcount\minutes 
\newcount\scratch 
 
\def\timestamp{%
\scratch=\time 
\divide\scratch by 60 
\edef\hours{\the\scratch} 
\multiply\scratch by 60 
\minutes=\time 
\advance\minutes by -\scratch 
---$\,$\hours:\null 
\ifnum\minutes< 10 0\fi 
\the\minutes} 

\title{Dominant next-to-leading order QCD corrections to Higgs plus three jet 
production in vector-boson fusion}

\author{Terrance Figy \\
Institute of Particle Physics Phenomenology,
Durham University, South Road, Durham, DH1 3LE, United Kingdom \\
E-mail: \email{terrance.figy@durham.ac.uk}}
\author{Vera Hankele$^a$ and Dieter Zeppenfeld$^b$\\
Institut f\"ur Theoretische Physik, 
Universit\"at Karlsruhe, P.O.Box 6980, 76128 Karlsruhe, Germany \\
E-mail:$^a$\email{vera@particle.uni-karlsruhe.de}, 
$^b$\email{dieter@particle.uni-karlsruhe.de}}
\abstract{
We present the calculation of the dominant next to leading order QCD 
corrections to Higgs boson production in association with three jets via 
vector boson fusion in the form of a NLO parton-level Monte Carlo program. 
QCD corrections to integrated cross sections are modest, while the 
shapes of some kinematical distributions change appreciably at NLO. Scale 
uncertainties are shown to be reduced at NLO for the total cross section 
and for distributions. We consider a central jet veto at the LHC and 
analyze the veto probability for typical vector boson fusion cuts.  
Scale uncertainties of the veto probability are 
sufficiently small at NLO for precise Higgs coupling measurements
at the LHC.
}

\keywords{Standard Model, Higgs Physics, QCD, NLO Computations}
\preprint{arXiv:0710.5621\\
          IPPP/07/82\\
          DCPT/07/164\\
          KA-TP-28-2007}
\begin{document}

%
\include{intro}
\include{details}

\include{results}

\include{conclusions}
%
%
%
\acknowledgments{
This research was supported in part by the Deutsche Forschungsgemeinschaft
under SFB TR-9 ``Computational Particle Physics'' and via the 
Graduiertenkolleg ``High Energy Physics and Particle Astrophysics'' and in part by the European Community's Marie-Curie Research
Training Network under contract MRTN-CT-2006-035505
`Tools and Precision Calculations for Physics Discoveries at Colliders'. We also thank the Galileo Galilei Institute for Theoretical
Physics for the hospitality and the INFN for partial support during the
completion of this work. }
%
%
\appendix
\include{appendix}

\include{ref}
\end{document}

%% file: intro.tex
\section{Introduction}
\label{sec:intro}
One of the primary goals of the CERN Large Hadron Collider (LHC) is the 
discovery of the Higgs boson and 
a thorough investigation of
the mechanism of electroweak (EW) symmetry breaking~\cite{ATLAS,CMS}. 
In this context, vector-boson fusion (VBF) has emerged as a
particularly interesting  class of processes. 
Higgs boson production in VBF, i.e.\ the EW
reaction $qq\,\to\, qqH$, where the Higgs decay products are detected in
association with two tagging jets, offers a promising
discovery channel~\cite{Rainwater:1999gg} and, once its existence
has been verified, will help to constrain the couplings of the Higgs
boson to gauge bosons and fermions~\cite{Zeppenfeld:2000td}. 

The observation of two forward tagging jets in Higgs production via 
VBF at the LHC is crucial for  the suppression of 
backgrounds~\cite{Kauer:2000hi,VBFhtautau,VBFhtoww,Rainwater:1999sd,
VBFhtophoton,Eboli:2000ze}.  In addition to forward jet tagging, the 
veto of any additional jet activity in the central region 
(\emph{central jet veto}) leads to further suppression of QCD backgrounds 
such as $W^{+}W^{-}jj$ , $t \bar{t}jj$, and  gluon fusion $Hjj$ 
production~\cite{Rainwater:1999sd,DelDuca:2004wt}.
This is due to the fact that the $t$-channel exchange of quarks or 
gluons tends to radiate harder and more central gluons
than in the VBF case. For VBF processes, jet activity in the central region 
is suppressed due to color singlet exchange in the $t$-channel.  
For the central jet veto (CJV) proposal, events are discarded if any 
additional jet with a transverse momentum above a minimal value, 
$p_{T,veto}$, is found between the tagging jets~\cite{Kauer:2000hi,Bjorken:1992er,Rainwater:1996ud,Barger:1994zq,Barger:1993qu,Barger:1991ar,Duff:1993ut}.

In order to utilize the CJV for the measurement of Higgs couplings, the  
reduction factor, $P_{\rm surv}$, caused by the CJV on the observable 
signal cross section must be precisely known. The relevant information 
is contained in the fraction of VBF Higgs events with at least one additional 
veto jet between the two tagging jets, i.e. we need to know the ratio of 
the 3-jet Higgs cross section, $\sigma_{jjj}$, to the inclusive cross 
section for VBF Higgs production with two tagging jets, $\sigma_{jj}$. 
The survival probability for the Higgs signal is then given by 
$P_{\rm surv}=1-\sigma_{jjj}/\sigma_{jj}$. Perturbative
survival probabilities for the CJV have been calculated for the Higgs 
boson signal and background processes using LO matrix elements 
\cite{Rainwater:1999sd,Rainwater:1996ud}.  The cross section for 
the VBF process $pp \rightarrow Hjjj$ is proportional to $\alpha_s$ at LO, 
which leads to substantial theoretical uncertainties (scale variations
of 30\% or more). Even though the effect on the survival probability 
is mitigated by the smallness of $\sigma_{jjj}/\sigma_{jj}$ (about 0.1 to 0.2
for veto thresholds $p_{T,veto}\approx 20$~GeV), a more reliable prediction
requires a calculation of the NLO QCD corrections to the $H jjj$ cross section.
We have performed this calculation and report on the results in this paper.
We do not consider additional reductions of the survival probability 
due to underlying event and pile-up effects. An assessment of these 
effects is best performed after first LHC data have become available.

A full NLO QCD calculation of the process $pp\to HjjjX$ involves
virtual corrections with hexagon diagrams and would be truly challenging.
As we explain in Sec.~\ref{sec:calc}, all pentagon and hexagon 
contributions are color suppressed by a factor $1/(N^2-1)$ in an SU(N) gauge
theory, and they are further suppressed by the kinematics of the VBF process.
For a prediction of the survival probability of the Higgs signal at the few
percent level, commensurate with the knowledge of the VBF cross section
for $Hjj$ production at NLO and expected experimental accuracies at the LHC, 
these contributions are completely negligible.
We therefore perform the calculation by systematically neglecting the gauge
invariant subsets of diagrams which involve $t$-channel gluon exchange and
which lead to pentagons and hexagons. Similarly, we neglect identical fermion
effects for four-quark final states. In Sec.~\ref{sec:calc} we more explicitly 
specify and justify these approximations and we briefly describe
the calculation of the LO and NLO matrix elements for $Hjjj$ production.

Section~\ref{sec:res} deals with phenomenological applications of the
parton-level Monte Carlo program which we have developed. We consider the
$Hjjj$ cross section at NLO after typical VBF cuts and discuss the 
reduction of the scale dependence of relevant distributions. We show that 
the scale dependence of $P_{surv}$ is reduced to about 1\% by including 
the NLO QCD corrections to the three jet cross section. Conclusions are
given in Sec.~\ref{sec:concl}.
Explicit formulas for the virtual corrections and for finite collinear 
terms from initial state radiation in gluon initiated processes are
collected in two Appendices.

%% file: details.tex
\FIGURE{
\epsfig{file=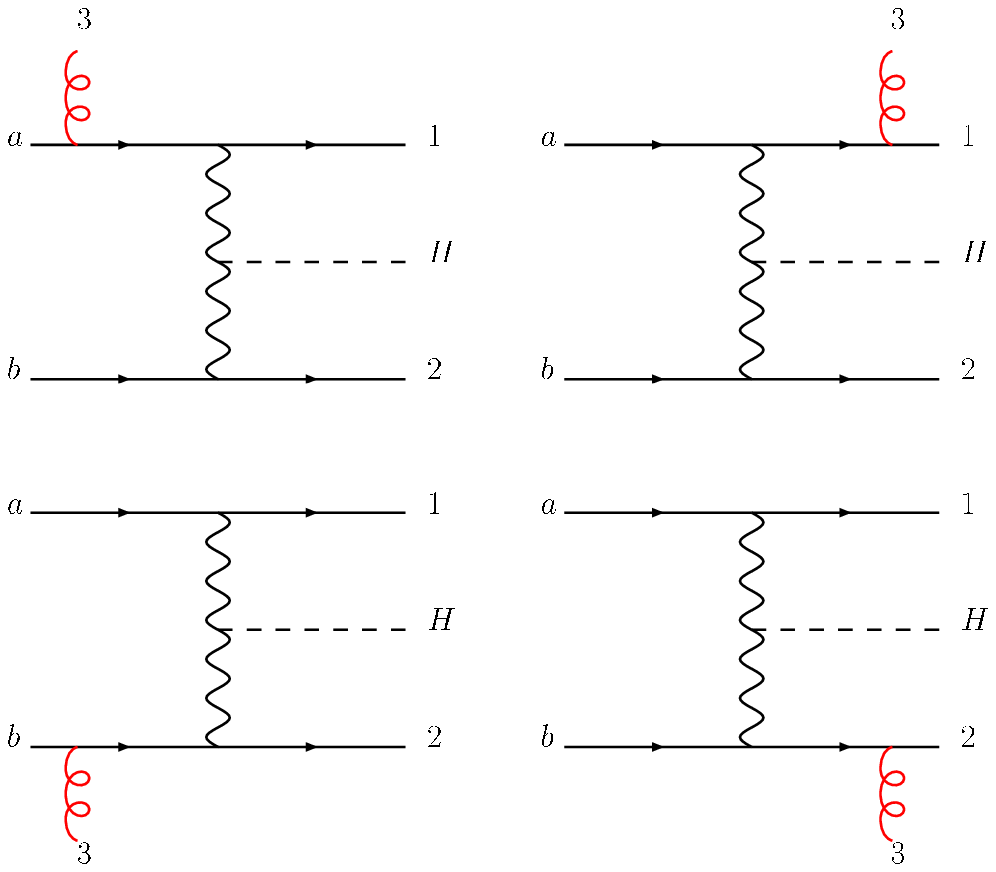,scale=0.8}
\caption{Lowest-order Feynman graphs for $pp \rightarrow Hjjj$ via VBF.}%
\label{fig:born1}}

\section{The NLO Calculation and Approximations}
\label{sec:calc}

The cross section for the leading order process $pp \to Hjjj$, via VBF, 
has been previously calculated as
the NLO real emission correction to $Hjj$ production in 
Refs.~\cite{Han:1992hr,Figy:2003nv,Berger:2004pca}. The relevant Feynman 
graphs are depicted in Fig.~\ref{fig:born1}:  one needs to consider 
the ${\cal O}(\alpha^3\alpha_s)$ subprocesses $qQ \to qQgH$ and crossed 
subprocesses with vector boson exchange in the $t$-channel. We explicitly
exclude $s$-channel weak boson exchange and thus set aside higgsstrahlung 
processes, i.e. $VHj$ production with subsequent decay $V\to jj$. In 
the following, higgsstrahlung is viewed as a separate process and we also
neglect any interference of VBF and higgsstrahlung (in the case of 
identical fermion flavors)
since these interference effects are very small in the phase 
space region relevant for VBF observation at the 
LHC~\cite{cogeorg,Ciccolini:2007jr}.
 
In order to clarify our notation and the approximations in our calculation, 
let us start by considering the form of the Born amplitude for the 
$qQ \to qQgH$ subprocess. By $\mc{M}_{3}(1_{q},2_{Q},3_{g},a_{q},b_{Q})$ 
we denote the matrix element for the parton level process
\beqn
q(p_{a}) + Q(p_{b}) \to q(p_{1}) + Q(p_{2}) + g(p_{3}) + H(P), 
\eeqn
shown in Fig.~\ref{fig:born1}. Two distinct color structures contribute
to this Born matrix element with three final state colored partons,
\beqn \label{eq:born1}
\mc{M}_{3}(1_{q},2_{Q},3_{g},a_{q},b_{Q})&=& 
\mc{A}_{3}(1_{q},3_{g},a_{q};2_{Q},b_{Q}) 
\delta_{i_{2} i_{b}} t_{i_{1} i_{a}}^{a_{3}} \nonumber \\
&+&  \mc{A}_{3}(2_{Q},3_{g},b_{Q};1_{q},a_{q}) 
\delta_{i_{1} i_{a}} t_{i_{2} i_{b}}^{a_{3}} \, .
\eeqn
Focusing on the gluon emission, each of the amplitudes $\mc{A}_{3}$ 
can be viewed as a Compton scattering amplitude for the process 
$Q(k_1)\to Q(k_2)g(q_1)V(q_2)$, defined by
\beqn
\mathcal{M}_{B}(k_2 ,q_1, q_2;\eps_1,\eps_2) &=& 
-e\,g_{\tau}^{VQ_{2}Q_{1}}g_{s}
\bar{\psi}(k_{2})\left \{ \gamma^{\nu}
\frac{(\slashit{k}_{2}+\slashit{q}_{2})}{(k_2+q_2)^2}\gamma^{\mu} \right.
\nonumber \\ 
&+& \left.  \gamma^{\mu}
\frac{(\slashit{k}_{2}+\slashit{q}_{1})}{(k_2+q_1)^2}\gamma^{\nu} \right \}
P_{\tau}\psi(k_{1})\epsilon_{1\mu}(q_1)\epsilon_{2\nu}(q_2)\; .
\label{eq:born_main}
\eeqn
Here, $-e\,g_{\tau}^{VQ_{2}Q_{1}}$ is the left- or righthanded coupling of
the quarks to the weak boson, $g_s$ denotes the strong coupling constant, 
$P_{\tau}=\frac{1}{2}(1+\tau \gamma^{5})$ is the chirality projector, and
$\epsilon_1$ and $\epsilon_2$ are the polarization vectors of the gluon and 
of the weak boson, respectively. The role of the polarization vector for the 
weak boson is taken by a current, $h^{\mu}$, which, for the first two
diagrams in Fig.~\ref{fig:born1}, is given by
\beqn \label{eq:hcur}
h^{\mu}(p_{b} \tau_{b},p_{2} \tau_{2}) = 
\delta_{\tau_{2} \tau_{b}} (-e) g_{HVV}
g_{\tau_{2}}^{Vf_{2}f_{b}} D_{V}[p_{a13}^{2}] D_{V}[p_{b2}^2] 
\bar{\psi}(p_{2}) \gamma^{\mu} P_{\tau_{2}} \psi(p_{b})
\eeqn
with $p_{ijk} = p_{i}-p_{j}-p_{k}$ and $p_{ij}=p_{i}-p_{j}$, while
$D_{V}[q^{2}]=1/[q^{2} - M_{V}^{2}]$ is the weak boson propagator, which,
in our calculation, only occurs with space-like momentum.
In terms of the Compton amplitude of Eq.~(\ref{eq:born_main}) the $\mc{A}_{3}$ 
are then given by
\beqn
\mc{A}_{3}(1_{q},3_{g},a_{q};2_{Q},b_{Q})=
\mc{M}_{B}(p_{1}, p_{3}, p_{a13}; \eps_{3}, h(p_{b} \tau_{b},p_{2} \tau_{2}))  ,
 \nonumber \\
\mc{A}_{3}(2_{Q},3_{g},b_{Q};1_{q},a_{q})=
\mc{M}_{B}(p_{2}, p_{3}, p_{b23}; \eps_{3},
h(p_{a} \tau_{a}, p_{1} \tau_{1}))  .
\eeqn

The $gQ \to q\bar{q}QH$ subprocess is obtained by crossing the
initial state quark $q(p_a)$ with the final state gluon in 
Eq.~(\ref{eq:born1}) and dropping the $s$-channel graphs which result from 
crossing the diagrams in the second line of Fig.~\ref{fig:born1}. 
The $3$-parton matrix elements $\mc{M}_{3}$ have been computed 
using the helicity amplitude method of Ref.~\cite{HZ}.  

The real emission corrections to VBF $Hjjj$ production consist of 
four subprocess 
classes with four final state partons.  These  classes are
(a) $q Q \rightarrow q Q ggH$, (b) $qQ \rightarrow qQ q'\bar{q}'H$, 
(c) $gQ \rightarrow q \bar{q}QgH$, and 
(d) $gg \rightarrow q \bar{q} Q \bar{Q}H$. The generalization to the crossed 
processes with $q \rightarrow \bar{q}$ and/or $Q \rightarrow \bar{Q}$ is 
straightforward.

\FIGURE{
\epsfig{file=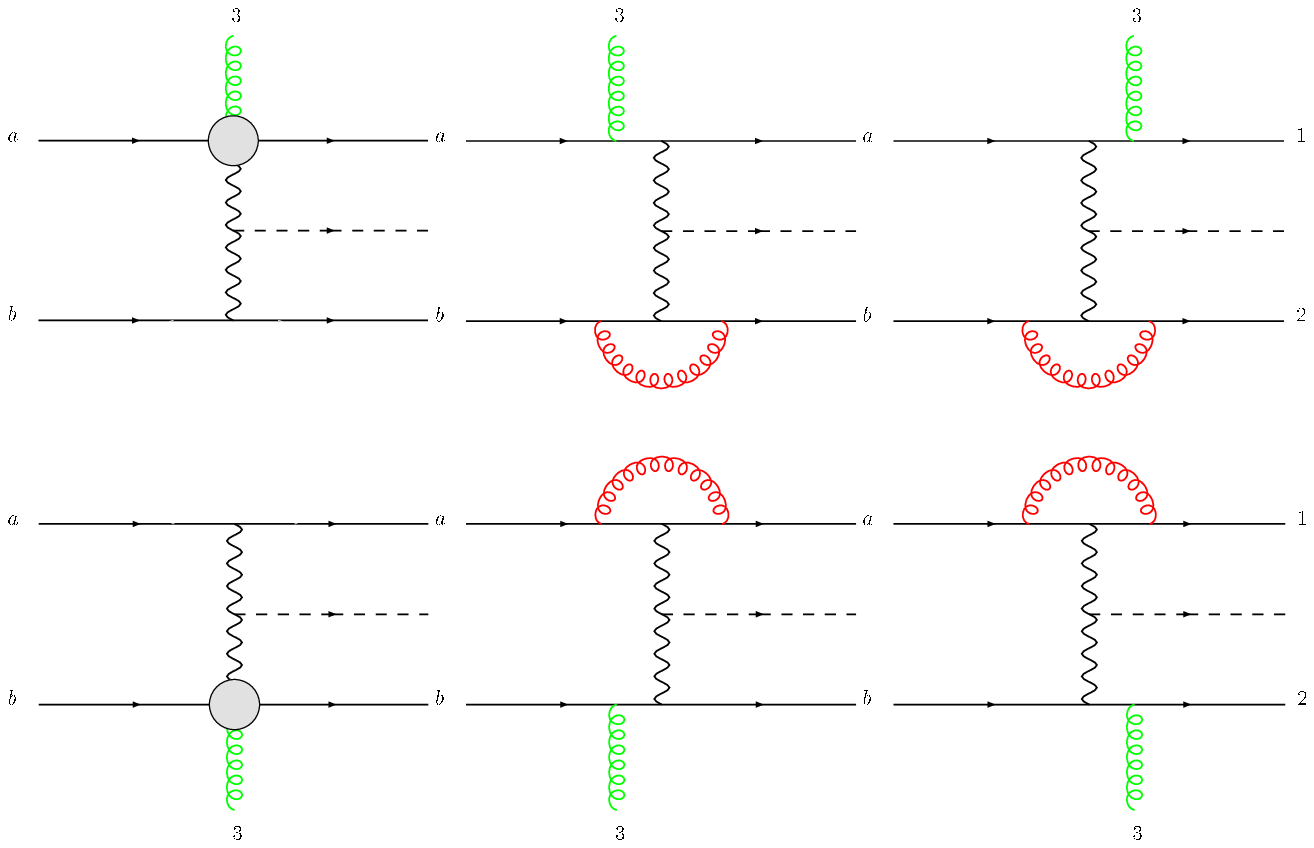,scale=0.9}
\caption{The dominant virtual QCD corrections. The ``blobs'' correspond to the
sum of all virtual corrections to the basic $Q\to QgV$ Compton amplitude and
are given more explicitly in Fig.~\ref{fig:bxln}. 
The first diagram and the second pair of diagrams in each line form 
gauge invariant subsets.}
\label{fig:box}}

The above subprocesses lead to soft and collinear singularities when 
integrated over the phase space of the final state partons. We use the 
Catani-Seymour dipole subtraction method to regulate these 
divergences~\cite{CS} and to cancel them against those originating 
from the virtual corrections. The virtual corrections can be divided into 
two classes of 
gauge covariant subsets. The first class (depicted in 
Fig.~\ref{fig:box}) are graphs in which the internal gluon propagator 
is attached to a single fermion line and which involve up to box
corrections.  The second class 
(depicted in Fig.~\ref{fig:hex}) are graphs in which the internal 
gluon propagates between different fermion lines, i.e. they contain 
a $t$-channel gluon. These graphs only play a role for subprocesses 
with two initial state quarks or anti-quarks. For gluon initiated processes
they only contribute to the interference of VBF and higgsstrahlung diagrams, 
which we neglect. The interference of the hexagon and pentagon 
amplitudes with the Born amplitude is 
color suppressed by a factor $d_{G}=N^2-1$ with respect to the interference 
of box corrections with the Born amplitude. 
We neglect the contribution of the hexagon and pentagon amplitudes.  
However, in doing so we must also consider the 
color structure of the real corrections and drop contributions which
cancel the infrared singularities of the pentagons and hexagons.

\FIGURE{
\epsfig{file=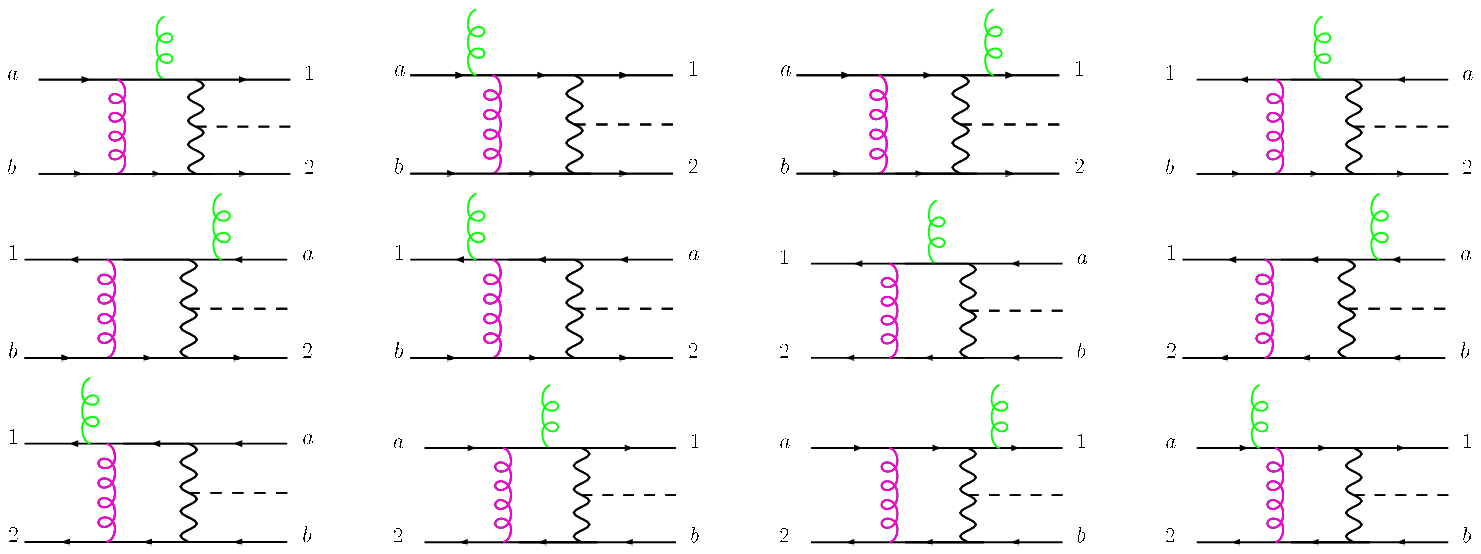,scale=0.9}
\caption{Pentagon and hexagon diagrams for the color structure 
$\delta_{i_{1} i_{a}} t_{i_{2} i_{b}}^{a_{3}}$. An analogous set appears 
with the external gluon attached to the lower quark line. Note that hexagon
graphs with a three-gluon-vertex correspond to a color structure which 
cannot interfere with the Born amplitude.}
\label{fig:hex}}

As an example for these real emission processes, 
consider the matrix element for the subprocess,
\beq
q(p_a ) + Q(p_b) \rightarrow 
q(p_1) + Q(p_2) + g(p_3) + g(p_4) + H(P)
\eeq
depicted in Fig.~\ref{fig:realcorr} and denoted by 
$ \mathcal{M}_{4}(1_{q},2_{Q},3_{g},4_{g},a_{q},b_{Q})$.
$ \mc{M}_{4}$ has the following color decomposition in terms of color 
subamplitudes, $\mc{A}$ and $\mc{B}$, 
\beqn\label{eq:real}
\mathcal{M}_{4}(1_{q},2_{Q},3_{g},4_{g},a_{q},b_{Q})
&=& (t^{a_3}t^{a_4})_{i_1 i_a} \delta_{i_2 i_b} 
\mathcal{A} (1_{q},3_{g},4_{g},a_{q};2_{Q},b_{Q})  \nonumber \\
&+& (t^{a_4}t^{a_3})_{i_1 i_a} \delta_{i_2 i_b} 
\mathcal{A} (1_{q},4_{g},3_{g},a_{q};2_{Q},b_{Q}) \nonumber \\
&+& (t^{a_3}t^{a_4})_{i_2 i_b} \delta_{i_1 i_a} 
\mathcal{A}(2_{Q},3_{g},4_{g},b_{Q};1_{q},a_{q})  \\
&+& (t^{a_4}t^{a_3})_{i_2 i_b} \delta_{i_1 i_a} 
\mathcal{A} (2_{Q},4_{g},3_{g},b_{Q};1_{q},a_{q}) \nonumber \\
&+&  t^{a_3}_{i_1 i_a} t^{a_4}_{i_2 i_b} 
\mathcal{B} (1_{q},3_{g},a_{q};2_{Q},4_{g},b_{Q}) \nonumber \\
&+&  t^{a_4}_{i_1 i_a} t^{a_3}_{i_2 i_b} 
\mathcal{B} (1_{q},4_{g},a_{q};2_{Q},3_{g},b_{Q}). \nonumber 
\eeqn
The $\mc{A}$ terms correspond to both gluons attached to the same
quark line, while $\mc{B}$ terms describe emission of one gluon from 
each of the two quark lines. Abbreviating these amplitudes by
\beqn\label{eq:abbrev}
\mathcal{A}_{1}&=& \mathcal{A}(1_{q},3_{g},4_{g},a_{q};2_{Q},b_{Q})\,, \qquad
\mathcal{A}_{2} =  \mathcal{A}(1_{q},4_{g},3_{g},a_{q};2_{Q},b_{Q})\,,
\nonumber \\
\mathcal{A}_{3}&=& \mathcal{A}(2_{Q},3_{g},4_{g},b_{Q};1_{q},a_{q})\,, \qquad
\mathcal{A}_{4} =  \mathcal{A}(2_{Q},4_{g},3_{g},b_{Q};1_{q},a_{q})\;,  \\
\mathcal{B}_{1}&=& \mathcal{B}(1_{q},3_{g},a_{q};2_{Q},4_{g},b_{Q})\,, \qquad
\mathcal{B}_{2} =  \mathcal{B}(1_{q},4_{g},a_{q};2_{Q},3_{g},b_{Q}), 
\nonumber 
\eeqn
the color summed squared matrix element can be written as
\beqn
|\mathcal{M}_{4}(1_{q},2_{Q},3_{g},4_{g},a_{q},b_{Q})|^2 &=&
d_{F}^2 C_{F}^2 \left \{ |\mathcal{A}_1|^2 + |\mathcal{A}_2|^2 +
  |\mathcal{A}_3|^2 + |\mathcal{A}_4|^2 + 
|\mathcal{B}_1|^2 + |\mathcal{B}_2|^2  \right. \nonumber \\
+  2x~\Re \left [  \mc{A}_1  \mc{A}_2^{*} +
 \mc{A}_3  \mc{A}_4^{*} \right ] 
&+& \left.  2y~\Re \left [ \left (\mc{A}_1 + \mc{A}_2 \right) 
\cdot   \left ( \mc{A}_3 + \mc{A}_4 \right )^{*} + 
    \mc{B}_1 \mc{B}_2^{*} \right ] \right \} 
\label{msq}
\eeqn
with $x=1-C_{A}/2C_{F}=-1/(N^2-1)$ and $y=1/d_{G}=1/(N^2-1)$, where the
explicit value is given for an $SU(N)$ gauge group, with $d_G = N^2-1$ 
and $d_F = N$. 
The term in Eq.~(\ref{msq}) which is proportional to $y$ leads to 
a soft divergence when integrated over the phase space of 
the soft/collinear parton, which is in fact canceled by 
the corresponding soft divergent hexagon and pentagon graphs. Since we neglect
the latter, for consistency, we also need to set $y=0$ in Eq.~(\ref{msq}). 
The association of the $y$-terms with the hexagon and pentagon diagrams of
Fig.~\ref{fig:hex} is made clear when recognizing that these are all
the contributions where the same gluon is attached to both an upper and 
a lower quark line. The $y$-term and the interference of hexagons and 
pentagons with the Born amplitude are not only color suppressed by a factor
$1/(N^2-1)$, they are further suppressed because the interfering amplitudes 
are never large simultaneously when typical VBF cuts are applied.
Consider, for example, the $\mc{B}_1$ and  $\mc{B}_2$ amplitudes in 
Fig.~\ref{fig:realcorr}. $\mc{B}_1$ is  large when $q_1$ and $g_3$ are
forward (i.e. in the initial $q_a$ direction) and $q_2$ and $g_4$ are 
backward, in the $q_b$ hemisphere. For $\mc{B}_2$ to be large, $q_1$ and 
$g_4$ must be forward while $q_2$ and $g_3$ are backwards.
These conditions cannot be satisfied simultaneously for a large rapidity 
separation between the highest $p_T$ jets, which typically will be the 
two quark jets. The largest interference between $\mc{B}_1$ and  $\mc{B}_2$ 
and, similarly, between $\mc{A}_1+\mc{A}_2$ and $\mc{A}_3+\mc{A}_4$ is to be 
expected when both factors in the interference terms have similar size,
i.e. when both gluons are emitted in the central region. For central
gluons, however, all contributing amplitudes are suppressed due to the gluon
radiation pattern of the underlying $t$-channel weak boson exchange.

\FIGURE[th]{
\epsfig{file=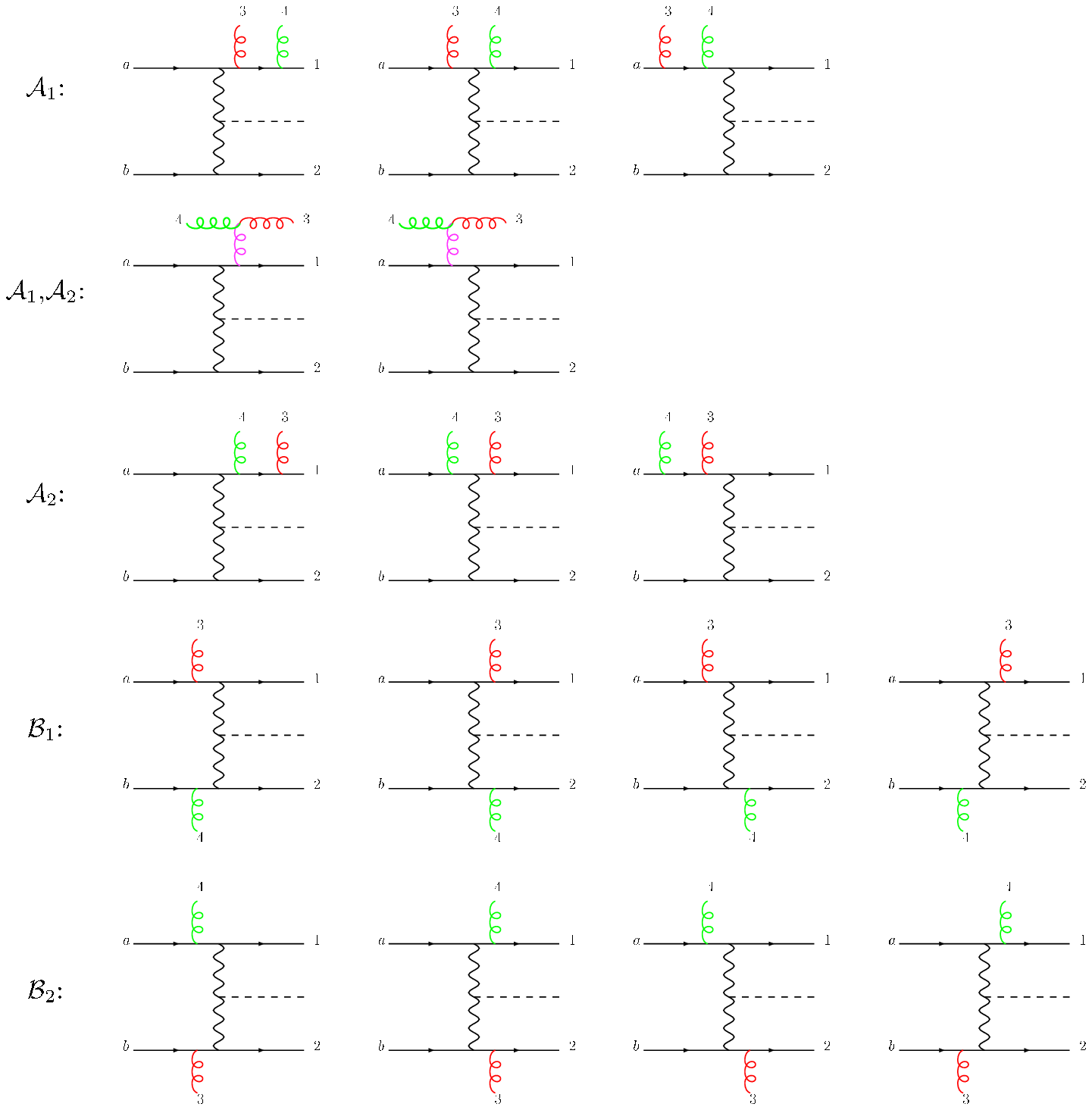,scale=0.9}
\caption{Feynman graphs for the real emission amplitude 
$\mathcal{M}_{4}(1_{q},2_{Q},3_{g},4_{g},a_{q},b_{Q})$ as described in
Eq.~(\ref{eq:real}) }
\label{fig:realcorr}}

We have estimated the error on the total $Hjjj$ cross section, 
$\Delta \sigma^{NLO}$, which we make in neglecting the hexagon and pentagon 
topologies (shown in Fig.~\ref{fig:hex}) and the corresponding interference
terms ($y$-terms) in Eq.~(\ref{msq}). Consider the dominant phase space region
where one gluon (say $g_3$) is hard and the second one ($g_4$) is soft. The 
soft emission can be factorized as an eikonal factor, while the hard part of
$\mc{B}_1$ and $\mc{A}_1+\mc{A}_2$ will be given by the upper line of the Born
diagram of Fig.~\ref{fig:born1}, i.e. by 
$\mc{A}_{3}(1_{q},3_{g},a_{q};2_{Q},b_{Q})$. Analogously, the hard factor in 
$\mc{B}_2$ and $\mc{A}_3+\mc{A}_4$ is 
$\mc{A}_{3}(2_{Q},3_{g},b_{Q};1_{q},a_{q})$, corresponding to hard emission
from the lower quark line in Fig.~\ref{fig:born1}. Approximately, in the soft
region, the $y$-terms plus the corresponding virtual corrections, given
by the interference of hexagons and pentagons with the Born amplitude,
are proportional to the product
\beq \label{eq:nloproxy}
\frac{\alpha_s}{2 \pi} \frac{d_{F}^2 C_{F}^2}{(N^2-1)} 2~  \Re \left [ 
\mc{A}_{3}(1_{q},3_{g},a_{q};2_{Q},b_{Q})
\mc{A}_{3}(2_{Q},3_{g},b_{Q};1_{q},a_{q})^*\right ]  \;,
\eeq
integrated over the 3-parton phase space. 
In Fig.~\ref{fig:int}, we compare the absolute value of this 
proxy for the full interference
terms (dotted blue curves) with the tree level cross section (dashed red
curve). Shown is the distribution in rapidity for the veto jet 
(lowest $p_T$ parton) measured with respect to the  
center of the rapidity of the two tagging jets. In the right panel of 
Fig.~\ref{fig:int}, the ratio of the two distributions is shown.
As a second estimator for the neglected terms, we have calculated the
full $y$-terms for the 4-parton final state and soft approximations for
the hexagons and pentagons, by keeping the infrared divergent
$C$-function terms only, according to the prescription of
Ref.~\cite{Dittmaier:2003bc}. For both contributions the full Catani-Seymour
subtraction has been implemented, with dipole terms as listed for
$y=1/d_G$ in Table~\ref{tbl:dip}. The resulting curve for 
\beqn
R(y_{\rm rel}) = \frac{d \Delta \sigma^{NLO}(\mu_R,\mu_F)/d y_{\rm rel}}
{d \sigma^{LO}(\mu_R,\mu_F)/dy_{\rm rel}} \; ,
\eeqn 
is shown for $\mu_R=\mu_F=40~{\rm GeV}$. The ratio, $R$, reaches a
maximal value of $ \approx 10^{-4}$ in the central region between the
two tagging jets, in agreement with the result for the proxy discussed
above. We conclude that the $y$-terms and the corresponding hexagon and
pentagon contributions give a relative contribution below one permille
everywhere in phase space and can safely be neglected.
We note that these interference terms are at the same level as the 
interference between gluon fusion and vector boson fusion 
for $pp \rightarrow Hjjj$. In a complete calculation, not only would the
hexagon and pentagon graphs need to be calculated, gluon fusion
contributions would have to be included as well. 

\FIGURE[th]{
\epsfig{file=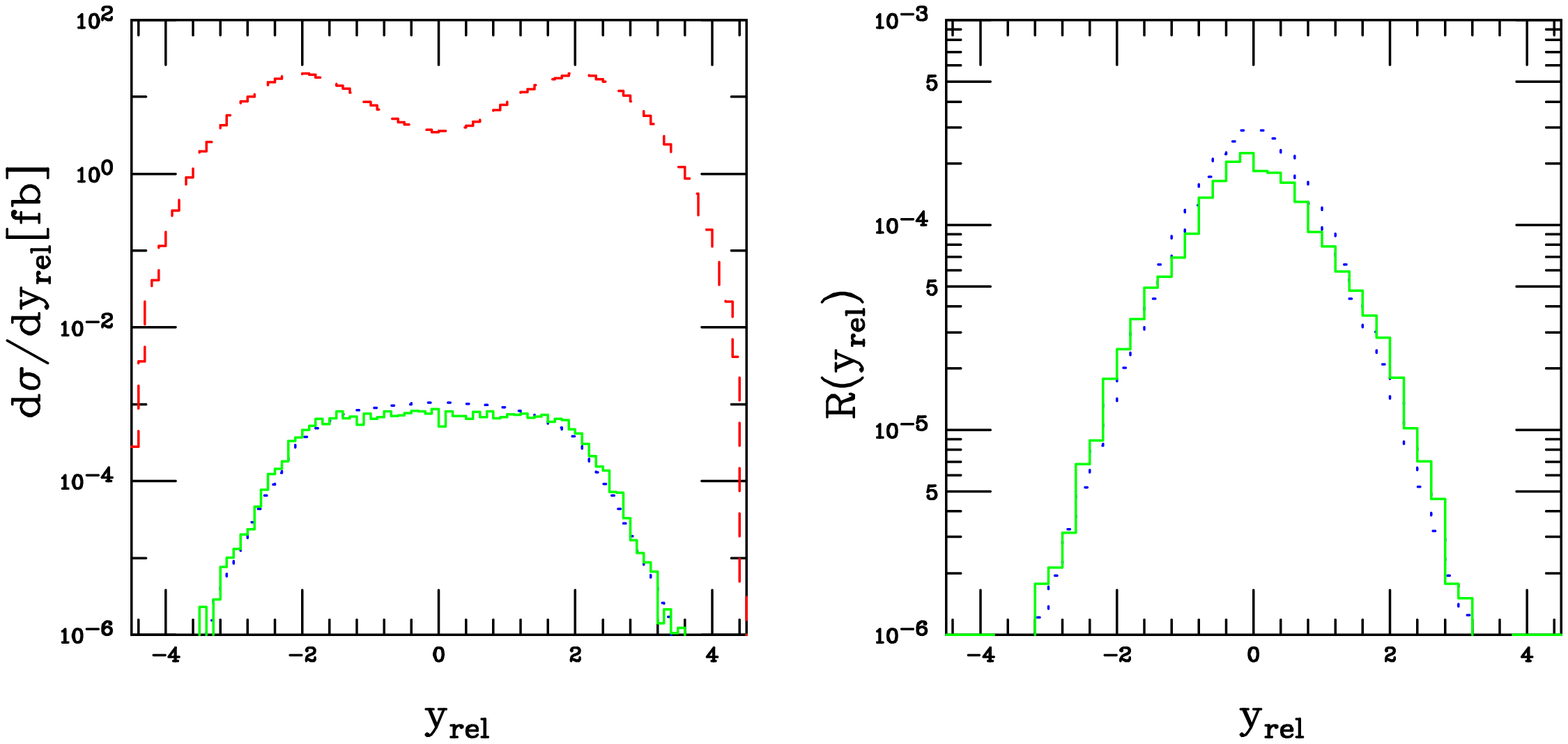,angle=90,width=5.0in}
\caption{The distribution in rapidity of the veto jet measured with 
respect to the center of rapidity of the tagging jets. In the left panel, 
$d\sigma/dy_{\rm rel}$ is shown at LO (dashed red), for the proxy given 
by Eq.~(\ref{eq:nloproxy}) (dotted blue), and for the NLO color suppressed 
contribution in the soft approximation (solid green).  The right panel 
depicts the ratio, $R$, for both the proxy (dotted blue) and the 
NLO color suppressed contribution (solid green). }
\label{fig:int}}

In addition to the $y$-terms discussed above we also neglect any 
interference terms for identical fermions in our simulations. These
terms are color suppressed by a factor $1/N$ and can only contribute when
fermion helicities are the same. For
charged current contributions we have determined the size of
these interference terms for 4-quark final states and have compared them
to the charged current contribution to the LO 3-jet cross section. 
We find a relative contribution of $7.5\cdot 10^{-4}$ within the cuts of
Section~\ref{sec:res}: also these ``Pauli interference terms'' are
truly negligible. With these approximations, the \fortran ~code for 
the real emission matrix element squared was generated with the help
of \MADGRAPH ~\cite{madgraph}. 

\TABLE{
\caption{Dipole factors for the real emission corrections to $Hjjj$ 
production. The $y=1/d_G$ line gives the additional dipole factors 
which are needed for the $qQ\to qQggH$ process when the $y$ terms 
in Eq.~(\protect\ref{msq}) are not neglected.} 
\begin{tabular}{|c|c|}
\hline
real subprocess & dipole factors \\
\hline
$(1_{q},2_{Q},3_{g},4_{g},a_{q},b_{Q})$ & $\mc{D}_{14,3}, \mc{D}_{13,4},\mc{D}_{34,1},\mc{D}_{14}^{a},\mc{D}_{13}^{a}$ \\
$y=0$ in Eq.~(\ref{msq}) & $\mc{D}^{a}_{34}, \mc{D}^{a4}_{1},\mc{D}^{a3}_{1},\mc{D}^{a4}_{3},\mc{D}^{a3}_{4}$\\
& $\mc{D}_{24,3}, \mc{D}_{23,4},\mc{D}_{34,2},\mc{D}_{24}^{b},\mc{D}_{23}^{b}$\\
& $\mc{D}^{b}_{34}, \mc{D}^{b4}_{2},\mc{D}^{b3}_{2},\mc{D}^{b4}_{3},\mc{D}^{b3}_{4}$\\
\hline
$y=1/d_{G}$ in Eq.~(\ref{msq}) & $\mc{D}^{a3,b},\mc{D}^{b3,a},\mc{D}^{a4,b},\mc{D}^{b4,a},\mc{D}^{a}_{24}$ \\
& $\mc{D}^{b}_{14},\mc{D}^{a4}_{2},\mc{D}^{b4}_{1},\mc{D}_{24,1},\mc{D}_{14,2}$ \\
& $\mc{D}^{a}_{23},\mc{D}^{b}_{13},\mc{D}^{a3}_{2},\mc{D}^{b3}_{1},\mc{D}_{23,1},\mc{D}_{13,2}$ \\
\hline
$(1_{q},2_{Q},3_{q'},4_{\bar{q}'},a_{q},b_{Q})$ & $\mc{D}_{34,1}, \mc{D}_{31,2}, \mc{D}^{a}_{34}, \mc{D}^{b}_{34}$\\
& $\mc{D}^{a1}_{3}, \mc{D}^{a1}_{4}, \mc{D}^{b2}_{3}, \mc{D}^{b2}_{4}$ \\
\hline
$(1_{q},2_{Q},3_{\bar{q}},4_{g},a_{g},b_{Q})$ & $\mc{D}^{a3}_{1},\mc{D}^{a1}_{3},\mc{D}^{a3}_{4},\mc{D}^{a1}_{4}$\\
& $\mc{D}^{a4}_{1},\mc{D}^{a4}_{3},\mc{D}^{b4}_{2},\mc{D}^{b}_{24}$\\
& $\mc{D}^{a}_{14},\mc{D}^{a}_{34},\mc{D}_{14,3},\mc{D}_{34,1}$\\
\hline
$(1_{q},2_{Q},3_{\bar{q}},4_{\bar{Q}},a_{g},b_{g})$ & $\mc{D}^{b4}_{2},\mc{D}^{b2}_{4},\mc{D}^{a3}_{1},\mc{D}^{a1}_{3}$ \\
\hline
\end{tabular}
\label{tbl:dip}}

The $4$-parton phase space integral of the squared real matrix elements 
suffers from 
soft and collinear divergences. The dipole subtraction method of Catani 
and Seymour provides a means to regulate these divergences \cite{CS}.
In the Catani-Seymour formalism the NLO corrections consist of three 
pieces: (a) the contribution of the dipole subtracted real corrections, 
$\sigma_{4}^{NLO}$, (b)  the contribution of the finite virtual 
corrections, $\sigma_{3}^{NLO}$, and (c) a piece resulting from the 
factorization of 
collinear singularities into the parton distribution functions, 
$\sigma^{NLO}_{3,\mathrm{col}}$.  As an example, consider the process, 
$qQ \to qQggH$, in the $y=0$ case in Eq.~(\ref{msq}). The subtracted 
cross section for this process takes the form,
\begin{eqnarray}\label{eq:sub}
\sigma_{4}^{NLO}(qQ\to qQggH) &=& 
\int_{0}^{1} dx_{a} \int_{0}^{1} dx_{b} f_{q/p}(x_{a},\mu_{F}) 
f_{Q/p}(x_{b},\mu_{F}) 
\frac{1}{2 \hat{s}} d\Phi_{5}(p_{a},p_{b}) \nonumber \\
&\cdot & \left \{ |\mc{M}_{4}(1_{q},2_{Q},3_{g},4_{g},a_{q},b_{Q})|^2 
F_{J}^{(4)}(\p1,\p2,\p3,\p4;p_{a}+p_{b}) \right. \nonumber \\
&-& \sum_{\mathrm{pairs}  \atop i,j} \sum_{k\neq i,j}  
\mc{D}_{ij,k}(\p1,\p2,\p3,\p4;p_{a},p_{b}) 
F_{J}^{(3)}( \p1,..\pt{ij},\pt{k},..\p4;\pa,\pb) \\
&-& \sum_{\mathrm{pairs}  \atop i,j}  \left [ 
\mc{D}_{ij}^{a}(\p1,\p2,\p3,\p4,;\pa,\pb) 
F_{J}^{(3)}(\p1,..\pt{ij},..,\p4;\pt{a},\p{b}) + 
(a \leftrightarrow b) \right] \nonumber \\
&-& \left . \sum_{i} \sum_{k \neq i}  \left[ 
\mc{D}_{k}^{ai}(\p1,\p2,\p3,\p3;\pa,\pb) 
F_{J}^{(3)}(\p1,..\pt{k},...,\p4;\pt{ai},\p{b})+
(a \leftrightarrow b) \right ] \right \},
\nonumber 
\end{eqnarray}
where the $\mc{D}_{ij,k}$ etc. are the dipole factors as defined in 
Ref.~\cite{CS}, $d\Phi_{5}$ is the 5-particle phase space measure and 
$\hat{s}=(\pa+\pb)^2$ denotes the center-of-mass energy. 
A complete list of the dipole factors in Eq.~(\ref{eq:sub}) is shown 
in Table~\ref{tbl:dip}.  Notice, that we do not need to consider dipole 
factors for which there is an initial state singularity with an 
initial state spectator for the case of $y=0$ because in this approximation
radiative corrections to the upper and the lower lines 
in Fig.~\ref{fig:born1} effectively decouple.
We also show in Table~\ref{tbl:dip} dipole factors for quark-gluon and 
gluon-gluon initiated processes. The functions $F_{J}^{(3)}$ 
and $F_{J}^{(4)}$ define the jet algorithm for $4$-parton and $3$-parton 
final states and must be infrared safe which formally means 
that $F_{J}^{(4)} \to F_{J}^{(3)}$ in any case where the $4$-parton and 
$3$-parton configurations are kinematically degenerate. 

The dipole factors are integrated in $d=4-2 \eps$ space-time dimension 
over the phase space of the soft/collinear parton.  Integrating the 
dipole factors for the processes, $qQ \to qQggH$ and $qQ \to qQ q' \bar{q'}H$, 
lead to the universal singular factor, $<\mathbf{I}(\eps)>$. 
For the parton level process 
\beqn
q(p_{a}) + Q(p_{b}) \to q(p_{1}) + Q(p_{2}) + g(p_{3}) + H(P) \, , 
\eeqn
we can split $<\mathbf{I}(\eps)>$ into two pieces according to,
\beqn
<\mathbf{I}(\eps)> = C_{F} \left( \mc{I}_{1}(\eps)+\mc{I}_{2}(\eps) \right)\, .
\eeqn
$\mc{I}_{1}(\eps)$ is a piece proportional to the Born-level color 
subamplitude squared, $|\mathcal{A}_{3}(1_{q},3_{g},a_{q};2_{Q},b_{Q})|^2$, 
and is  
\begin{eqnarray}\label{eq:ieps1}
\mc{I}_{1}(\epsilon)&=&
|\mathcal{A}_{3}(1_{q},3_{g},a_{q};2_{Q},b_{Q})|^2 
\frac{\alpha_{s}(\mu_{R}^2)}{2 \pi} \frac{1}
{\Gamma(1-\epsilon)} \\ \nonumber 
&\cdot&
\left \{ \frac{1}{2} \left ( 
\left ( \frac{4 \pi \mu_{R}^2}{s_{13} }\right)^{\epsilon}
+ \left ( \frac{4 \pi \mu_{R}^2}{s_{a3} } \right )^{\epsilon} \right) 
\left( \frac{C_{A}}{\eps^2}+\frac{\gamma_{g}}{\eps} + \gamma_{g}+K_{g} 
\right) \right. \\ \nonumber 
&+& \frac{1}{2} \frac{C_A}{C_F} \left ( \left (\frac{4 \pi \mu_{R}^2}{s_{13}} 
\right)^{\epsilon}
+ \left (\frac{4 \pi \mu_{R}^2}{s_{a3}} \right)^{\epsilon}
-2\left (\frac{4 \pi \mu_{R}^2}{s_{a1}} \right)^{\epsilon} \right )
\left(\frac{C_{F}}{\eps^2}+\frac{\gamma_{q}}{\eps}+
 \gamma_{q}+K_{q}  \right)  \\ \nonumber 
&+&2 \left ( \left ( \frac{4 \pi \mu_{R}^2}{s_{a1} }\right)^{\epsilon}
+\left.  \left ( \frac{4 \pi \mu_{R}^2}{s_{b2} } \right )^{\epsilon} \right) 
\left(\frac{C_{F}}{\eps^2}+\frac{\gamma_{q}}{\eps}+
\gamma_{q}+K_{q}  \right)  \right \}  \, 
\end{eqnarray}
with
\beq
\gamma_{q} = \frac{3}{2} C_{F}, \quad 
\gamma_{g} = \frac{11}{6} C_{A} - \frac{2}{3} T_{R} N_{f},
\eeq 
and, 
\beq
K_{q} = \left ( \frac{7}{2} - \frac{\pi^2}{6} \right ) C_{F}, \quad 
K_{g} = \left (\frac{67}{18} - \frac{\pi^2}{6} \right ) 
C_{A} - \frac{10}{9} T_{R} N_{f} \, .
\eeq
Here $s_{ij}=2 p_{i} \cdot p_{j}$ with $i=1,2,3,a$ or $b$. 
$T_{R}=1/2$, $C_{A}=N$, and $C_{F}=(N^2-1)/(2N)$ in $SU(N)$ 
gauge theory. The number of flavors is $N_{f}=5$. $\mc{I}_{2}(\eps)$ is 
obtained from $\mc{I}_{1}(\eps)$ by interchanging the quark labels,
$a \leftrightarrow b$ and 
$1 \leftrightarrow 2$. The $1/\eps^2$ and $1/\eps$ divergences cancel 
against the virtual corrections shown in Fig.~\ref{fig:box}.

In our approximation there are two distinct color structures that 
contribute to this virtual matrix element, 
$\mc{M}_{3}^{\mathrm{virt}}(1_{q},2_{Q},3_{g},a_{q},b_{Q})$, 
with three final state colored partons, 
\beqn \label{eq:virt}
\mc{M}_{3}^{\mathrm{virt}}(1_{q},2_{Q},3_{g},a_{q},b_{Q})&=& 
\mc{A}_{3}^{\mathrm{virt}}(1_{q},3_{g},a_{q};2_{Q},b_{Q}) 
\delta_{i_{2} i_{b}} t_{i_{1} i_{a}}^{a_{3}} \\ \nonumber
&+&  \mc{A}_{3}^{\mathrm{virt}}(2_{Q},3_{g},b_{Q};1_{q},a_{q}) 
\delta_{i_{1} i_{a}} t_{i_{2} i_{b}}^{a_{3}} \, .
\eeqn
The interference between the virtual and Born three parton amplitudes 
takes on the following form upon summing over
final state colors and averaging over initial state colors,
\beqn \label{eq:brvirt}
\overline{\sum_{\rm colors}}
2~\Re[\mc{M}_{3}^{\mathrm{virt}} \mc{M}_{3}^{*}] &=& 
C_{F}\left( 2~\Re[\mc{A}_{3}^{\mathrm{virt}}(1_{q},3_{g},a_{q};2_{Q},b_{Q}) 
 \mc{A}_{3}^{*}(1_{q},3_{g},a_{q};2_{Q},b_{Q})] \right.  \\ \nonumber
&+& \quad \quad \left. 
2~\Re[\mc{A}_{3}^{\mathrm{virt}}(2_{Q},3_{g},b_{Q};1_{q},a_{q}) 
 \mc{A}_{3}^{*}(2_{Q},3_{g},b_{Q};1_{q},a_{q}) ] \right ) \, .
\eeqn
We split the virtual corrections shown in Fig.~\ref{fig:box} into two 
classes: the virtual corrections along a quark line with only one weak
boson attached and the virtual corrections along a quark line with a gluon 
and a weak boson attached.
The former, with only a weak boson vertex,
are factorizable in terms of the tree-level current $h^{\mu}$ defined by 
Eq.~(\ref{eq:hcur}). For vertex corrections to the lower line one has
\beqn
h^{\mu}_{\mathrm{virt}}(\pb \tau_{b},\p2 \tau_{2})= 
h^{\mu}(\pb \tau_{b},\p2 \tau_{2}) C_F \frac{\alpha_s(\mu_R)}{4 \pi} 
\left(\frac{4 \pi \mu_R^2}{s_{b2}} \right)^{\epsilon} 
\frac{1}{\Gamma(1-\epsilon)} 
\left( -\frac{2}{\epsilon^2}-\frac{3}{\epsilon}  - 8 \right) \, ,
\eeqn
and similarly for vertex corrections to the upper line.
Here $\mu_{R}$ is the renormalization scale, and $s_{b2}=2 \pb \cdot \p{2}$ 
is the weak boson virtuality for massless quarks.

\FIGURE[t]{
\epsfig{file=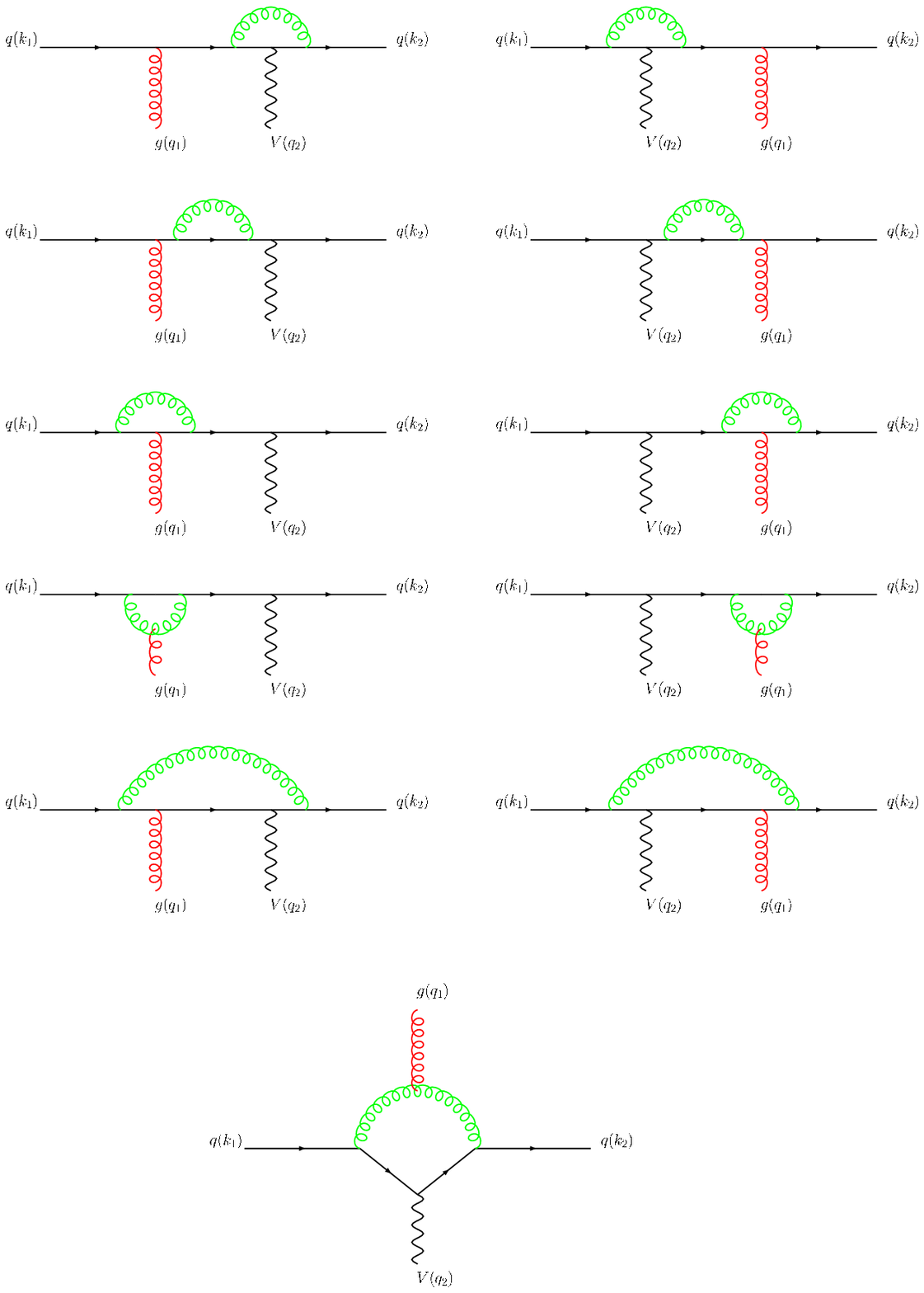,scale=0.9}
\caption{The one-loop QCD corrections for $q \to qgV$.}%
\label{fig:bxln}}

The second class of diagrams, corresponding to the ``blob'' in 
Fig.~\ref{fig:box} and shown explicitly in Fig.~\ref{fig:bxln}, are the 
virtual QCD corrections to the Feynman graphs where
a gluon $g$ and an electroweak boson $V$ ( outgoing momenta $q_1$ and $q_2$) 
are attached to the same fermion line.  The kinematics is given by
\beqn
Q(k_1) \rightarrow Q(k_2)+g(q_1)+V(q_2),
\eeqn
where $k_{1}^{2}=k_{2}^{2}=q_1^2=0$ and momentum conservation reads 
$k_1=k_2+q_2+q_1$.  As in Ref.~\cite{Oleari:2003tc}, it is 
convenient to use the Mandelstam variables for a $2\rightarrow 2$ process 
which is taken to be $\bar{q}q \rightarrow gV$.
The Mandelstam variables are thus defined as
\beqn \label{eq:mns}
s=(k_1 - k_2)^2 = (q_1 + q_2)^2, \nonumber \\
t = (k_1 - q_1)^2=(k_2+q_2)^2, \\ \nonumber 
u = (k_1 - q_2)^2=(k_2+q_1)^2. 
\eeqn
The gluon polarization denoted by $\eps_1(q_1)$ is transverse, i.e. 
$\eps_1 \cdot q_1 =0$ and this permits simplifications in the virtual
amplitudes.  The electroweak boson $V$ is 
always virtual in the calculation. Its effective polarization vector 
$\eps_2^\mu (q_2)$ corresponds to the tree level fermion current $h^\mu$. 
Due to the emission of the Higgs boson off the $t$-channel vector boson
propagator, this fermion current is not conserved.  
Hence, terms with $\eps_2 \cdot q_{2}$ must be kept.  However, 
electroweak gauge invariance of the amplitude is preserved, 
i.e., $\mathcal{M}_{\mu} q_{2}^{\mu} =0$.  We have computed the virtual 
amplitude in the conventional dimensional regularization scheme (CDR) and 
used Passarino-Veltman reduction in $d=4-2 \epsilon$ spacetime dimensions to 
reduce tensor loop integrals into scalar loop 
integrals~\cite{Passarino:1978jh}. The virtual amplitude 
$\mathcal{M}_{V} =\mathcal{M}_{V}(k_2,q_1 ,q_2;\eps_1,\eps_2)$ for 
$Q(k_1) \rightarrow Q(k_2) g(q_1) V(q_2)$ is
\beqn \label{eq:boxline}
\mathcal{M}_{V} &=&\mathcal{M}_{B} \frac{\alpha_s(\mu_{R}^2)}{4 \pi} 
\frac{1}{\Gamma(1-\epsilon)}
\left \{ \frac{1}{2} 
\left(\left (\frac{4 \pi \mu_{R}^2}{-u }\right)^{\epsilon} +
\left (\frac{4 \pi \mu_{R}^2}{-t}\right)^{\epsilon} \right) 
(-\frac{C_A}{\epsilon^2}-\frac{\gamma_{g}}{\epsilon}) \right. \nonumber \\
&+& \frac{1}{2} \frac{C_A}{C_F} \left ( \left( \frac{4 \pi \mu_{R}^2}{-u} 
\right)^{\epsilon} +  
\left( \frac{4 \pi \mu_{R}^2}{-t} \right)^{\epsilon}-2  
\left( \frac{4 \pi \mu_{R}^2}{-s} \right)^{\epsilon} \right)
(-\frac{C_F}{\epsilon^2}-\frac{\gamma_q}{\epsilon}) \nonumber \\
&+&2 \left (\frac{4 \pi \mu_{R}^2}{-s} \right)^{\epsilon}
\left. (-\frac{C_F}{\epsilon^2} -\frac{\gamma_q}{\epsilon}) 
+F(-s,-t,-u) -\frac{\pi^2}{6} C_{A}- 8 C_{F}  \right \} \\ \nonumber 
&+& \widetilde{\mc{M}}_{V} \, ,
\eeqn
where 
\beqn \label{eq:Fstu}
F(-s,-t,-u) = \frac{C_A}{2} \left( \ln^2 \left(\frac{-u }{\mu_{R}^2} \right)  
+\ln^2 \left(\frac{-t }{\mu_{R}^2} \right)\right)
-\frac{1}{2}(C_A - 2 C_F) \ln^2 \left( \frac{-s }{\mu_{R}^2} \right) \nonumber \\
+\frac{3}{2} (C_A - 2 C_F) \ln \left( \frac{-s }{\mu_{R}^2} \right) 
+ (\frac{1}{3} T_{R} N_{f}-\frac{5}{3}C_A)
\left( \ln \left(\frac{-u }{\mu_{R}^2}\right) + 
\ln \left(\frac{-t }{\mu_{R}^2}\right) \right)  \, .
\eeqn
The finite part 
$\widetilde{\mc{M}}_{V}=\widetilde{\mc{M}}_{V}(k_2,q_1 ,q_2;\eps_1,\eps_2)$ 
is given by
\beqn \label{eq:finvirt}
 \widetilde{\mc{M}}_{V}&=&\frac{\alpha_s(\mu_{R}^2)}{4 \pi} 
(-e) g_{\tau}^{V Q_2 Q_1} g_s \\ \nonumber 
 &\cdot &  \left \{(C_F - \frac{1}{2}C_A)
\{\widetilde{\mathcal{M}}^{(1)}_{\tau}(k_2,q_1,q_2;\eps_1,\eps_2) + 
\widetilde{\mathcal{M}}^{(2)}_{\tau}(k_2,q_1,q_2;\eps_1,\eps_2) \}\right.\\ 
\nonumber
&-& \left. \frac{1}{2}C_A  
\widetilde{\mathcal{M}}^{(3)}_{\tau}(k_2,q_1,q_2;\eps_1,\eps_2) \right \} \, .
\eeqn
Results for physical kinematic regions can be obtained through the analytic continuation of  
Eq.~(\ref{eq:boxline}) by the replacement of the time-like invariant by $s \rightarrow s+i0^{+}$, 
$t \rightarrow t+i0^{+}$, or $u\rightarrow u+i0^{+}$.  The $i\pi$ factors which result from the 
analytic continuation vanish upon interfering the virtual amplitude with the Born amplitude. The 
analytic continuation for any double logarithms is dealt with automatically by the \fortran~code for the
finite part of the virtual amplitude $\widetilde{\mc{M}}_{V}$ given by Eq.~(\ref{eq:finvirt}).

The $\widetilde{\mc{M}}_{\tau}^{(i)}$ for $i=1,2,3$ are finite and can be 
expressed in terms of the finite parts of the Passarino-Veltman, $B_{0}$, 
$C_{0}$, and $D_{ij}$ functions, which we denote as 
$\widetilde{B}_{0}$, $\widetilde{C}_{0}$, and $\widetilde{D}_{ij}$. 
Analytic expressions for the 
$\widetilde{\mc{M}}_{\tau}^{(i)}$ are given in Appendix~\ref{app:A}.

We can build the virtual color subamplitudes of Eq.~(\ref{eq:virt}) out of 
the two classes of virtual corrections discussed above.
The virtual color subamplitudes are then
\beqn
\mc{A}_{3}^{\mathrm{virt}}(1_{q},3_{g},a_{q};2_{Q},b_{Q}) &=&\mc{M}_{B}
(\p1,\p3,p_{a13};\eps_{3},h_{\mathrm{virt}}(\pb \tau_{b},\p2 \tau_{2})) 
\nonumber \\
&+&\mc{M}_{V}(\p1,\p3,p_{a13};\eps_{3},h(\pb \tau_{b},\p2 \tau_{2})) \, , \\
\mc{A}_{3}^{\mathrm{virt}}(2_{Q},3_{g},b_{Q};1_{q},a_{q}) &=&\mc{M}_{B}
(\p2,\p3,p_{b23};\eps_{3},h_{\mathrm{virt}}(\pa \tau_{a},\p1 \tau_{1})) 
\nonumber \\
&+&\mc{M}_{V}(\p2,\p3,p_{b23};\eps_{3},h(\pa \tau_{a},\p1 \tau_{1})).
\eeqn 
For the following we adopt the following abbreviations,
\beqn
\mc{A}_{3,1a}&=&\mc{A}_{3}(1_{q},3_{g},a_{q};2_{Q},b_{Q}), \quad \quad 
\mc{A}_{3,2b}= \mc{A}_{3}(2_{Q},3_{g},b_{Q};1_{q},a_{q}) \\ \nonumber 
\mc{A}_{3,1a}^{\mathrm{virt}}&=&
\mc{A}_{3}^{\mathrm{virt}}(1_{q},3_{g},a_{q};2_{Q},b_{Q}), \quad \quad 
\mc{A}_{3,2b}^{\mathrm{virt}}= 
\mc{A}_{3}^{\mathrm{virt}}(2_{Q},3_{g},b_{Q};1_{q},a_{q}). 
\eeqn
The color decomposed interference of the Born and virtual subamplitudes 
is then,
\beqn \label{eq:box1}
 \overline{\sum_{\rm colors}}\;  2~\Re[\mc{A}_{3,1a}^{\mathrm{virt}} 
 \mc{A}_{3,1a}^{*}]  
 &=&|\mathcal{A}_{3,1a}|^2 \frac{\alpha_{s}(\mu_{R}^2)}{2 \pi} \frac{1}
{\Gamma(1-\epsilon)} \nonumber \\
&\cdot &
\left \{ \frac{1}{2} \left ( \left ( \frac{4 \pi \mu_{R}^2}{s_{13} }
\right)^{\epsilon}
+ \left ( \frac{4 \pi \mu_{R}^2}{s_{a3} } \right )^{\epsilon} \right) 
(-\frac{C_A}{\epsilon^2} 
-\frac{\gamma_{g}}{\epsilon})
 \right.  \nonumber \\ 
&+& \frac{1}{2} \frac{C_A}{C_F} \left ( \left (\frac{4 \pi \mu_{R}^2}{s_{23}} 
\right)^{\epsilon}
+ \left (\frac{4 \pi \mu_{R}^2}{s_{a3}} \right)^{\epsilon}
-2\left (\frac{4 \pi \mu_{R}^2}{s_{a1}} \right)^{\epsilon} \right )
(-\frac{C_F}{\epsilon^2} 
-\frac{\gamma_{q}}{\epsilon}) 
  \nonumber \\
&+& 2 \left ( \left ( \frac{4 \pi \mu_{R}^2}{s_{a1} }\right)^{\epsilon}
+  \left ( \frac{4 \pi \mu_{R}^2}{s_{b2} } \right )^{\epsilon} \right)
(-\frac{C_F}{\epsilon^2} 
-\frac{\gamma_{q}}{\epsilon}) \nonumber \\ 
&-& \left. \frac{\pi^2}{6} C_{A} -16 C_{F} + F(s_{a1},s_{a3},s_{13}) 
\right \}  \nonumber \\
&+& 2~\Re[\widetilde{\mc{A}}_{3}^{\mathrm{virt}}
(1_{q},3_{g},a_{q};2_{Q},b_{Q}) 
 \mc{A}_{3,1a}^{*}]   \, ,
\eeqn
with $\widetilde{\mc{A}}_{3}^{\mathrm{virt}}(1_{q},3_{g},a_{q};2_{Q},b_{Q})
=\widetilde{\mc{M}}_{V}(\p1,\p3,p_{a13};\eps_{3},
h(\pb \tau_{b},\p2 \tau_{2}))$. A similar expression for 
$2~\Re[\mc{A}_{3,2b}^{virt} \mc{A}_{3,2b}^{*}]$ is obtained by making 
the replacements, 
$a \leftrightarrow b$ and $1 \leftrightarrow 2$, in Eq.~(\ref{eq:box1}). 

Summing together the contributions from Eq.~(\ref{eq:ieps1}) and 
Eq.~(\ref{eq:box1}) yields the finite $3$-parton NLO cross section
\beqn \label{eq:sig1}
\sigma^{NLO}_{3}(q Q \rightarrow q Q g H) &=& 
\int_{0}^{1} dx_{a} \int_{0}^{1} dx_{b} f_{q/p}(x_a,\mu_F) f_{Q/p}(x_b,\mu_F) 
\nonumber \\ 
&\times &\frac{1}{2 \hat{s}} d \Phi_{4}(\pa,\pb) 
F_{J}^{(3)}(\p1,\p2,\p3,P;\pa,\pb)  \\ 
&\cdot &\left \{ |\mathcal{M}_{B}(1_{q},2_{Q},3_{g},a_{q},b_{Q})|^2 
\left ( 1+ \frac{\alpha_{s}(\mu_{R}^2)}{2 \pi} K_{\rm born}\right)  \right. 
\nonumber \\ 
&+&|\mathcal{A}_{3}(1_{q},3_{g},a_{q};2_{Q},b_{Q})|^2 \frac{\alpha_{s}(\mu_{R}^2) 
C_{F}}{2 \pi} F(s_{a1},s_{a3},s_{13}) \nonumber \\ 
&+&|\mathcal{A}_{3}(2_{Q},3_{g},b_{Q};1_{q},a_{q})|^2 \frac{\alpha_{s}(\mu_{R}^2) 
C_{F}}{2 \pi} F(s_{b2},s_{b3},s_{23}) \nonumber \\ 
&+& C_{F}\left( 2~\Re[\widetilde{\mc{A}}_{3}^{\mathrm{virt}}
(1_{q},3_{g},a_{q};2_{Q},b_{Q}) 
 \mc{A}_{3}^{*}(1_{q},3_{g},a_{q};2_{Q},b_{Q})] \right. \nonumber \\
&+& \left. \left. 2~\Re[\widetilde{\mc{A}}_{3}^{\mathrm{virt}}
(2_{Q},3_{g},b_{Q};1_{q},a_{q}) 
 \mc{A}_{3}^{*}(2_{Q},3_{g},b_{Q};1_{q},a_{q}) ] \right ) \right \}, 
\nonumber 
\eeqn
with 
\beq
K_{\rm born} = \left (-\frac{2 \pi^2}{3} + \frac{50}{9} \right ) C_A - 
\frac{16}{9} T_{R} N_{f}
+ 2 C_{F} \left ( 2-\pi^2 \right ).
\eeq

The remaining divergent piece of the integral of the dipole factors 
in Eq.~(\ref{eq:sub}) is proportional to the $P^{qq}$ and $P^{gq}$ 
splitting functions and is factorized into the parton distribution 
functions. The surviving finite collinear terms are given by
\beqn \nonumber 
\sigma_{3,{\rm col}}^{NLO}(qQ \rightarrow qQ gH) &=& \int_{0}^{1} dx_{a} 
\int_{0}^{1} dx_{b} \frac{1}{2 \hat{s}} 
 d \Phi_{4}(\pa,\pb)
F_{J}^{(3)}(\p1,\p2,\p3;\pa,\pb) \\ \nonumber 
& \cdot & \{ ( f_{q/p}(x_{a},\mu_F) f_{Q/p}^{2,b}(x_{b},\mu_F,\mu_R) + 
f_{q/p}^{1,a}(x_{a};\mu_F,\mu_R) f_{Q/p}(x_{b};\mu_F))
\\ \nonumber  
&\cdot & |\mc{M}_{3}(1_{q},2_{Q},3_{g};a_{q},b_{Q})|^2 \\ \nonumber 
&+& \frac{1}{2} C_{A} f_{q/p}(x_{a};\mu_F) ( f_{Q/p}^{3,b}(x_{b};\mu_F,\mu_R) 
- f_{Q/p}^{2,b}(x_{b};\mu_F,\mu_R)) 
\\ \nonumber 
&\cdot &|\mc{A}_{3}(2_{Q},3_{g},b_{Q};1_{q},a_{q})|^2 \\ 
&+& \frac{1}{2} C_{A} (f_{q/p}^{3,b}(x_{a};\mu_F,\mu_R) - 
f_{q/p}^{1,a}(x_{a};\mu_F,\mu_R)) f_{Q/p}(x_{b},\mu_F) 
\\ \nonumber 
& \cdot & | \mc{A}_{3}(1_{q},3_{g},a_{q};2_{Q},b_{Q}) |^2 \},
\eeqn
and similarly for the anti-quark initiated processes. Here the quark 
functions $f_{q/p}^{i,j}(x;\mu_{F},\mu_{R})$ are given by 
\begin{eqnarray}
f^{i,j}_{q/p}(x;\mu_F,\mu_R)&=&\frac{\alpha_s(\mu_R)}{2 \pi} \int_{x}^{1} 
\frac{dz}{z}
\left\{ f_{g/p}\left(\frac{x}{z};\mu_F \right) A^{i,j}_{gq}(z) \right. 
\nonumber \\ 
&+&\left [f_{q/p} \left (\frac{x}{z};\mu_F \right) - z f_{q/p}(x;\mu_F) 
\right] B^{i,j}_{qq}(z)  \\ \nonumber 
&+&\left.  f_{q/p}\left(\frac{x}{z};\mu_F \right) C^{i,j}_{qq}(z) \right \}
+ \frac{\alpha_s(\mu_R)}{2 \pi} f_{q/p}(x;\mu_F) D^{i,j}_{qq}(x) 
\nonumber \, , 
\end{eqnarray}
with kernels
\begin{eqnarray}
A^{i,j}_{gq}(z)&=& T_R [z^2+(1-z)^2]\ln \frac{2p_j p_i(1-z)}{\mu_F^2 z}
+T_R 2z(1-z) \, . \\
B^{i,j}_{qq}(z)&=&C_F \left [\frac{2}{1-z} \ln \frac{2p_j p_i(1-z)}{\mu_F^2}-
\frac{\gamma_i}{C_i}\frac{1}{1-z} \right]  , \\
C^{i,j}_{qq}(z)&=&C_F \left [-(1+z)\ln \frac{2p_j p_i(1-z)}{\mu_F^2 z} -
\frac{2}{1-z} \ln z + (1-z) \right], \\
D^{i,j}_{qq}(x)&=&C_F \left [\frac{2 \pi^2}{3} - 5 - \frac{\gamma_i}{C_i}-
\frac{\gamma_i}{C_i} \ln(1-x)+\ln^2(1-x) \right. \\
&+&\frac{3}{2}\ln\frac{2p_i p_j}{\mu_F^2}+
\left. 2 \ln(1-x)\ln\frac{2p_i p_j}{\mu_F^2} \right ] .\nonumber 
\end{eqnarray}
where if parton $i$ is a gluon, $C_{i}=C_{A}$ and if parton $i$ is a quark or anti-quark, 
$C_{i}=C_{F}$.  Likewise, $\gamma_{i}=\gamma_{q}$ if parton $i$ is a quark or anti-quark and
$\gamma_{i}=\gamma_{g}$ if parton $i$ is a gluon. 
The analogous results for gluon initiated processes are 
given in Appendix~\ref{app:B}.

We have implemented the QCD corrections for $pp \rightarrow Hjjj$ into a 
fully flexible parton-level Monte Carlo program. We have checked 
the dipole subtraction by verifying that the dipole subtraction terms 
and the real emission matrix elements match in the various singular 
regions.  The gauge invariance of the virtual matrix elements has been 
checked numerically for random choices of momenta. The finite collinear 
counter-terms that remain after the factorization of initial-state 
collinear divergences have been obtained by two independent calculations. 
We have also introduced a cut, $\alpha \in (0,1]$, on the phase space of the 
dipoles as described in Ref.~\cite{Nagy:2003tz}. We have checked that the integrated 
cross section is independent of this parameter and have used $\alpha=0.3$ in
our simulations.

In all subsequent calculations we use the input parameters for defining 
Standard Model (SM) couplings as listed in Table \ref{tbl:smparm}.  
Other SM couplings are computed using LO
electroweak relations. Cross sections are computed using CTEQ6M parton
distributions \cite{cteq6} for all NLO results and CTEQ6L1 parton
distributions for all leading order cross sections.  The running of the
strong coupling is evaluated at two-loop order, with
$\alpha_s(M_Z)=0.118$, for all NLO results.  For LO results, the running
of the strong coupling is evaluated at one-loop with
$\alpha_s(M_Z)=0.130$. In order to reconstruct jets from the final-state
partons, the $k_T$ algorithm \cite{kToriginal} as described in
Ref.~\cite{kTrunII} is used, with resolution parameter $D=0.8$. 

\TABLE[t]{
\caption{Standard Model input parameters}
\begin{tabular}{cccccc}
\hline
\hline
$\alpha_{s}^{NLO}(M_Z)$ & $\alpha_{s}^{LO}(M_Z)$ & $M_{Z}$  &  $M_{W}$ & 
$G_{F}$ & $m_{h}$\\
 $0.118$ &  $0.130$ & $91.188~{\rm GeV}$ & $80.416~{\rm GeV}$ & 
$1.16639 \times 10^{-5}/{\rm GeV}^2$ & $120 ~{\rm GeV}$ \\ 
\hline
\hline
\end{tabular}%
\label{tbl:smparm}}

%% file: results.tex
\section{Predictions for the LHC}
\label{sec:res}
The goal of our calculation is a precise prediction of the LHC cross
section for Higgs boson production in VBF with three or more jets. The
$k_T$ algorithm is used to define jets and these jets are required to have 
\begin{eqnarray}
\label{cuts1}
p_{Tj} \geq 20~{\rm GeV} \, , \qquad\qquad |y_j| \leq 4.5 \, .
\end{eqnarray}
Here $y_j$ denotes the rapidity of the (massive) jet momentum which is 
reconstructed as the four-vector sum of massless partons of 
pseudorapidity $|\eta|<5$. 

At LO, there are exactly three massless final state partons. 
At NLO these jets may be
composed of two partons (recombination effect) or four well-separated
partons may be encountered, of which at least three 
satisfy the cuts of Eq.~(\ref{cuts1}) and would give rise to either  
three or four-jet events. As with LHC data, a choice needs
to be made for selecting the tagging jets in such a multijet situation.
Here the ``$p_{T}$-method" is chosen.  For a given event, the tagging
jets are defined as the two jets with the highest transverse momentum
with 
\beq\label{eq:tagcuts}
p_{Tj}^{\rm tag} \ge 30 ~{\rm GeV}, \quad \quad |y_{j}^{\rm tag}| \le 4.5.
\eeq
The non-tagging jets by default are jets of lowest 
transverse momenta. They do not need to satisfy the cuts of 
Eq.~(\ref{eq:tagcuts}) but must satisfy the cuts of Eq.~(\ref{cuts1}). 

The Higgs boson decay
products (generically called ``leptons" in the following) are required
to fall between the two tagging jets  
in rapidity and they should be well observable. While the exact criteria
for the Higgs decay products will depend on the channel considered, such
specific requirements here are substituted by  
generating isotropic Higgs boson decay into two massless ``leptons"
(which represent $\tau^{+} \tau^{-}$ or $\gamma \gamma$ final states) and requiring  
\begin{eqnarray}
p_{T\ell} \geq 20~{\rm GeV} \,,\qquad |\eta_{\ell}| \leq 2.5  \,,\qquad 
\triangle R_{j\ell} \geq 0.6 \, ,\end{eqnarray} 
where $\triangle R_{j\ell}$ denotes the jet-lepton separation in the
rapidity-azimuthal angle plane. In addition, the two ``leptons" are
required to fall between the two tagging jets in rapidity: 
\begin{eqnarray}
y_{j,min}^{\rm tag}+0.6<\eta_{\ell_{1,2}}<y_{j,max}^{\rm tag}-0.6.
\end{eqnarray} 
Note that no reduction due to branching ratios for specific final states
has been included in the calculation.  

Backgrounds to vector-boson fusion are significantly suppressed by
requiring a large rapidity separation for the two tagging jets. 
Tagging jets are required to reside in opposite detector hemispheres with
\beq \label{eq:ophem}
y_{j}^{\rm tag ~1} \cdot y_{j}^{\rm tag ~2} < 0
\eeq
and to have a large rapidity separation of 
\begin{eqnarray}\label{eq:yjjcut}
\Delta y_{jj} = | y_{j}^{\rm tag ~1}-y_{j}^{\rm tag ~2}|>4\;,
\end{eqnarray}
sometimes called ``rapidity gap cut''. 
QCD backgrounds for the Higgs signal typically occur at small
invariant masses, due to a the dominance of gluons at small Feynman $x$
in the incoming protons \cite{Rainwater:1999sd}. 
The QCD backgrounds can be reduced by  imposing a
 lower bound on the invariant mass of the tagging jets of 
\begin{eqnarray} \label{eq:dijet}
 m_{jj} =\sqrt{ (p_{j}^{\rm tag~1}+p_{j}^{\rm tag~2})^2} > 600 ~{\rm GeV}.
\end{eqnarray}

The cross section for Higgs production via VBF in association with
three jets or more ($Hjjj$), within the cuts of
Eqs.~(\ref{cuts1})-(\ref{eq:dijet}), 
is shown in Fig.~ \ref{fig:sigmaxi}.  The scale dependence of the NLO
and LO cross sections is shown for 
factorization and renormalization scales, $\mu_F$ and $\mu_R$, which are
tied to a fixed reference scale $\mu_0=40~{\rm GeV}$,
\beq
\mu_{R} = \xi_{R} \mu_{0}, \quad \quad \mu_{F} = \xi_{F} \mu_{0}.
\eeq
The value $\mu_0=40~{\rm GeV}$ was chosen to minimize the scale
dependence of the NLO predictions and at the same time it provides
optimal agreement of the LO approximation with the NLO result.

\FIGURE{
\epsfig{file=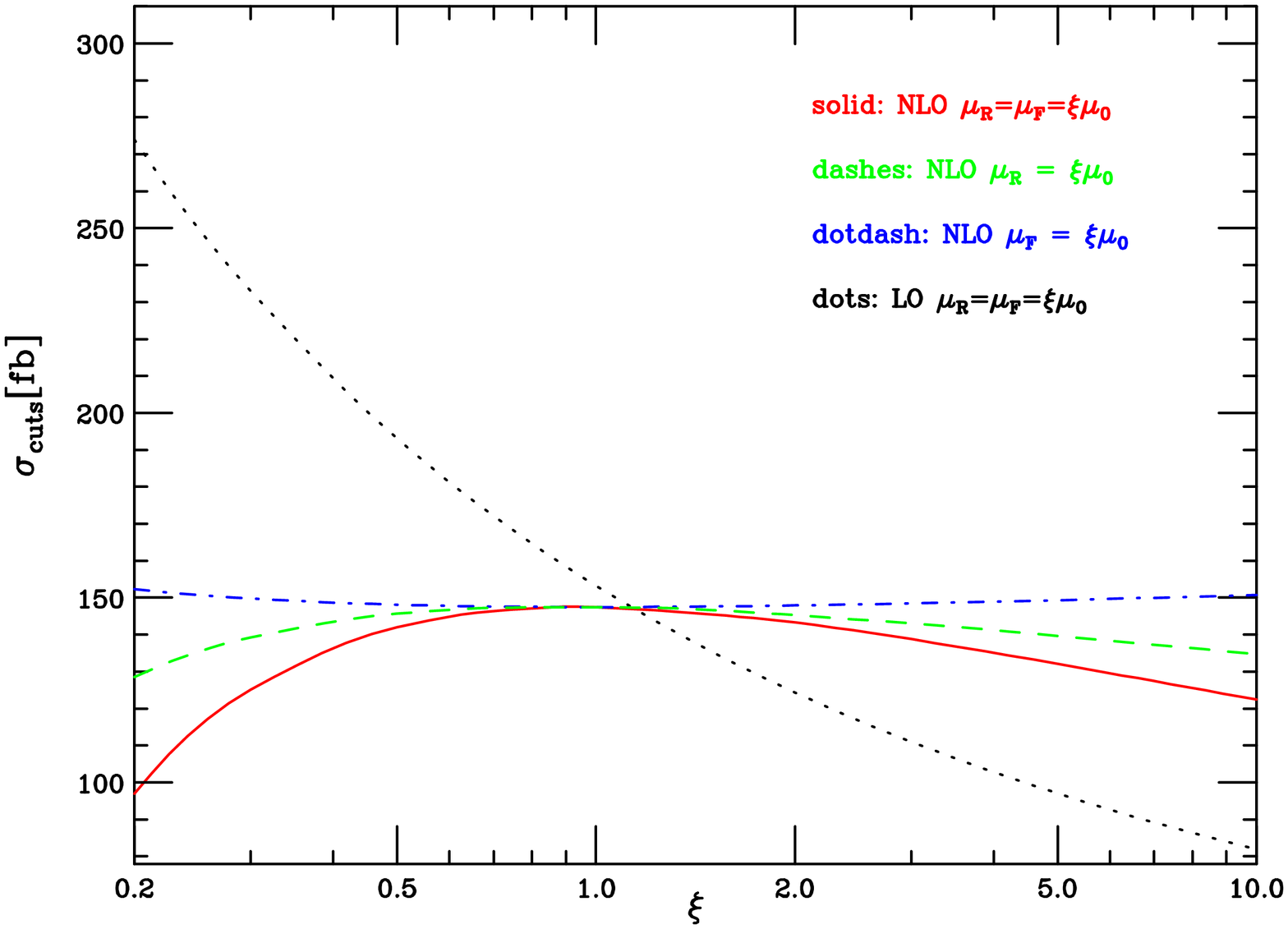,angle=90,width=4.5in}
\caption{Scale dependence of the total cross section at LO and NLO within 
the cuts of Eqs. (\ref{cuts1})-(\ref{eq:dijet}) for VBF $Hjjj$ production
at the LHC. The factorization scale $\mu_F$ and the renormalization scale 
$\mu_R$ are taken as multiples, $\xi \mu_{0}$,  of the fixed reference 
scale $\mu_{0} = 40~{\rm GeV}$. The NLO curves are  
for $\mu_R=\mu_F=\xi \mu_{0}$ (solid red line), 
$\mu_F = \mu_{0}$ and $\mu_R=\xi \mu_{0}$ (dashed green line),
and $\mu_F=\xi \mu_{0}$ and $\mu_R = \xi \mu_{0}$ (dot-dashed blue line ).  
The dotted black curve shows the scale dependence of the LO cross section 
for $\mu_R = \mu_F = \xi \mu_0$.}%
\label{fig:sigmaxi}}

The LO cross section depends on both the factorization and
renormalization scale.  For $\mu_{R} = \mu_{F}=\xi \mu_{0}$ with
$0.5<\xi<2$ the scale variation is  
$+26\%$ to $-19\%$ for the LO cross section.  The large scale variation
is primarily due the fact that the LO $Hjjj$ production cross section is
proportional to $\alpha_{s}$. This is in contrast to $Hjj$ production in
VBF, which only depends on the factorization scale at LO. 
At NLO three choices are shown: (a) $\xi_{R}=\xi_{F}= \xi$ (solid red line);  
(b) $\xi_{R} = \xi$, $\xi_{F}=1$ (dashed green line); (c) $\xi_{R} = 1$,
$\xi_{F} = \xi$ (dot-dashed blue line).  Allowing for a factor $2$
variation in either direction, 
i.e., considering the range $0.5\le \xi \le 2$, the NLO cross section
changes by less than $5\%$ in all cases.   

\FIGURE{
\epsfig{file=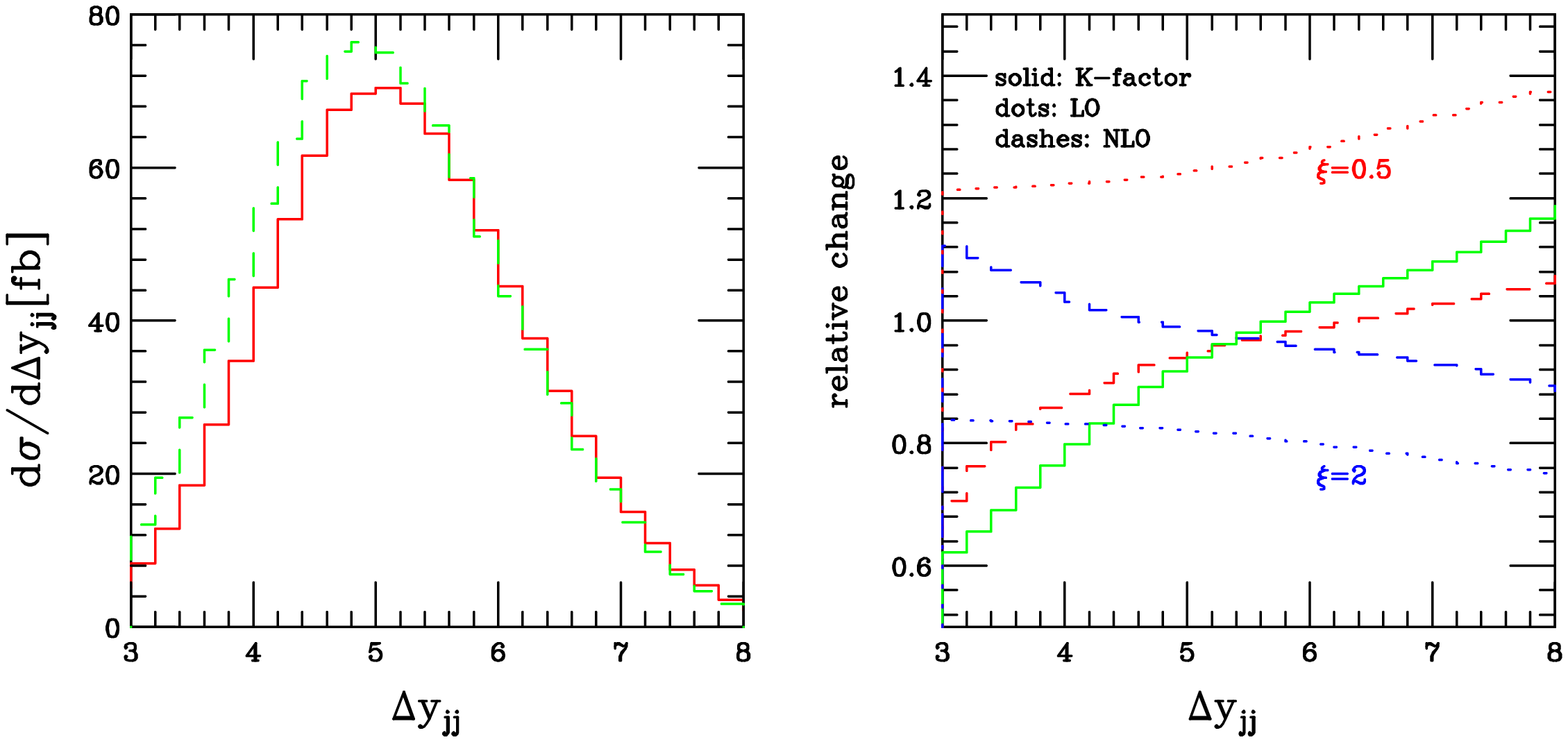,angle=90,width=5.5in}
\caption{Rapidity separation in $Hjjj$ production within the cuts of 
Eqs.~(\ref{cuts1})-(\ref{eq:ophem}) and Eq.~(\ref{eq:dijet}).  
In the left panel, $d\sigma/d \Delta y_{jj}$ is shown at LO (dashed green)
and NLO (solid red) for $\mu_F = \mu_R =\mu_0=40~{\rm GeV}$.  The right-hand 
panel depicts the $K$ factor (solid green) and scale variations of LO 
(dotted) and NLO (dashed) results for $\mu_R =\mu_F= \xi \mu_{0}$ 
with $\xi = 1/2$ and $2$.}%
\label{dsdyjjrel}}

Our Monte Carlo program allows the analysis of arbitrary infrared and collinear
safe distributions with NLO QCD accuracy. In order to assess the impact of 
the NLO corrections we compare LO and NLO results by plotting the dynamical 
$K$ factor
\beq
K(x) = \frac{d \sigma_{3}^{NLO}(\mu_{R} = \mu_{F}=\mu_{0})/dx}
{d \sigma_{3}^{LO}(\mu_R=\mu_F = \mu_{0})/dx}
\eeq
for our fixed reference scale of $\mu_0=40$~GeV. The stability of the results 
is represented via the scale dependence, given by the ratio of cross sections
and dubbed ``relative change'' in the following,
\beq
{\rm relative ~change} = \frac{d \sigma_{3}(\mu_R=\mu_F=\xi \mu_{0})/dx}
{d \sigma_{3} (\mu_R=\mu_F=\mu_{0})/dx}\;.
\eeq 
We plot results for $\xi = 1/2$ and $2$ with $\mu_{0}=40~{\rm GeV}$ for 
NLO and LO distributions.

\FIGURE{
\epsfig{file=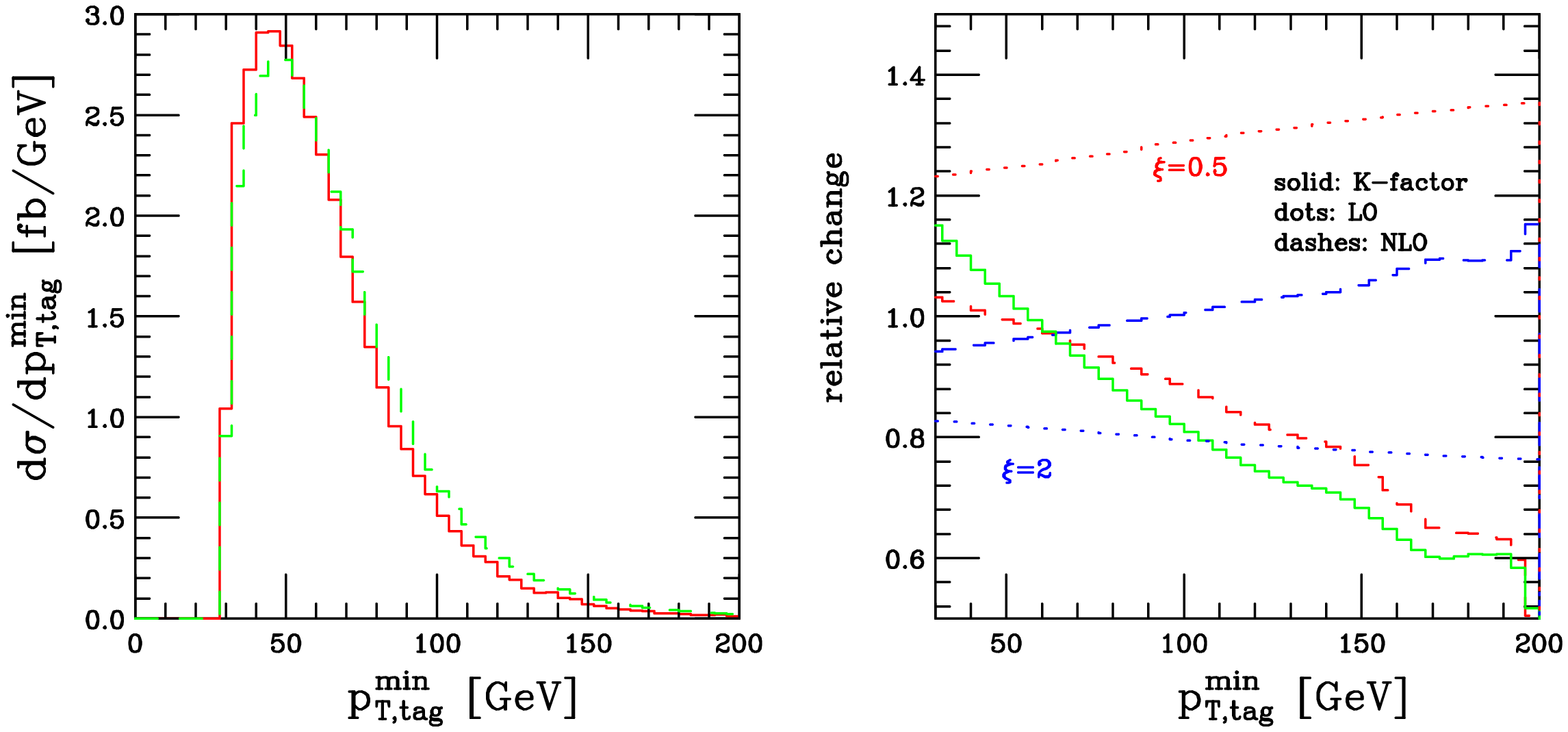,angle=90,width=5.5in}
\caption{Transverse momentum distribution for the softer tagging jet in
$Hjjj$ production within the cuts of Eqs.~(\ref{cuts1})-(\ref{eq:dijet}).  
The meaning of the curves is the same as in Fig.~\ref{dsdyjjrel}.}%
\label{dsdptmin}}

The wide separation in rapidity of the tagging jets is a characteristic 
feature of VBF processes. In the left-hand panel of Fig.~\ref{dsdyjjrel} 
the distribution $d\sigma/ d \Delta y_{jj}$ is shown at LO (dashed green) 
and at NLO (solid red) for $Hjjj$ production.  Just as in the NLO $Hjj$ 
case~\cite{Figy:2003nv}, the NLO corrections push the peak 
towards higher values of rapidity separation $\Delta y_{jj}$. 
This strengthens the case for the rapidity gap cut of $\Delta y_{jj}>4$.  
The $K$ factor (solid green) in the right-hand-side of Fig.~\ref{dsdyjjrel} 
is strongly phase space dependent. The scale variations $\xi=2^{\pm1}$ are 
significantly reduced by the NLO corrections, from $\approx 25\%$ at LO to 
$\approx 10\%$ or less at NLO in the relevant region $4<\Delta y_{jj}<7$.
Similar results are found for the transverse momentum distribution of the 
tagging jets as shown in Fig.~\ref{dsdptmin}.

\FIGURE{
\epsfig{file=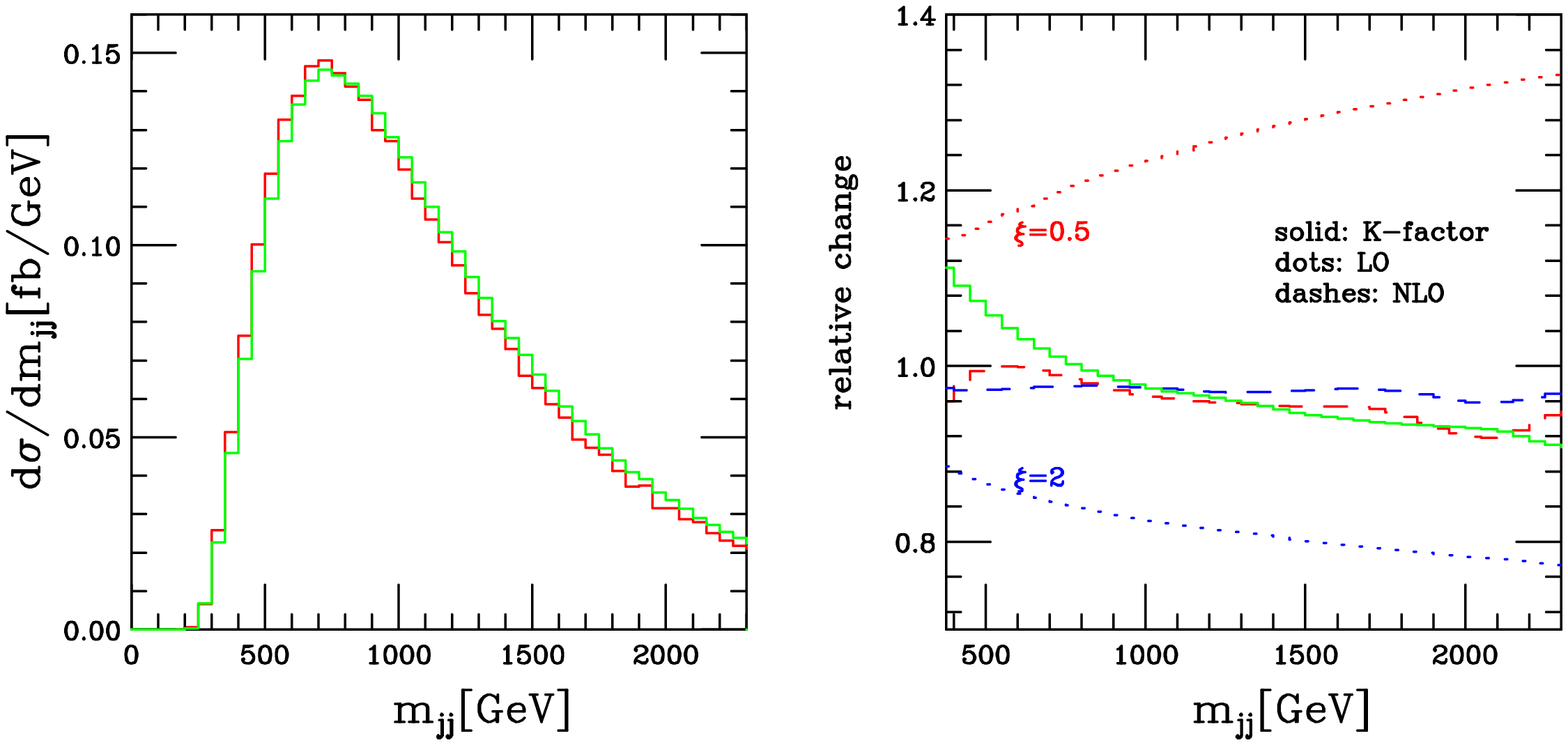,angle=90,width=5.5in}
\caption{Invariant mass distribution of the two tagging jets in 
$Hjjj$ production within the cuts of Eqs.~(\ref{cuts1})-(\ref{eq:ophem}) 
and Eq.~(\ref{eq:yjjcut}).  
The meaning of the curves is the same as in Fig.~\ref{dsdyjjrel}.}%
\label{dsdmjjrel}}

In Fig.~\ref{dsdmjjrel} the invariant mass distribution of the two 
tagging jets is shown for a fixed value of renormalization and factorization 
scale, $\mu_R = \mu_F =40~{\rm GeV}$. The $K$ factor (solid green) deviates 
from unity by 10\% or less for this distribution and this scale choice, i.e.
the LO result provides for an excellent estimate. The $\xi = 2^{\pm1}$ 
scale variations produce changes in the LO distribution of about 30\%,
however (dotted lines). This uncertainty is reduced to the 5\% level at 
NLO (dashed curves). 

%
%
\FIGURE{
\epsfig{file=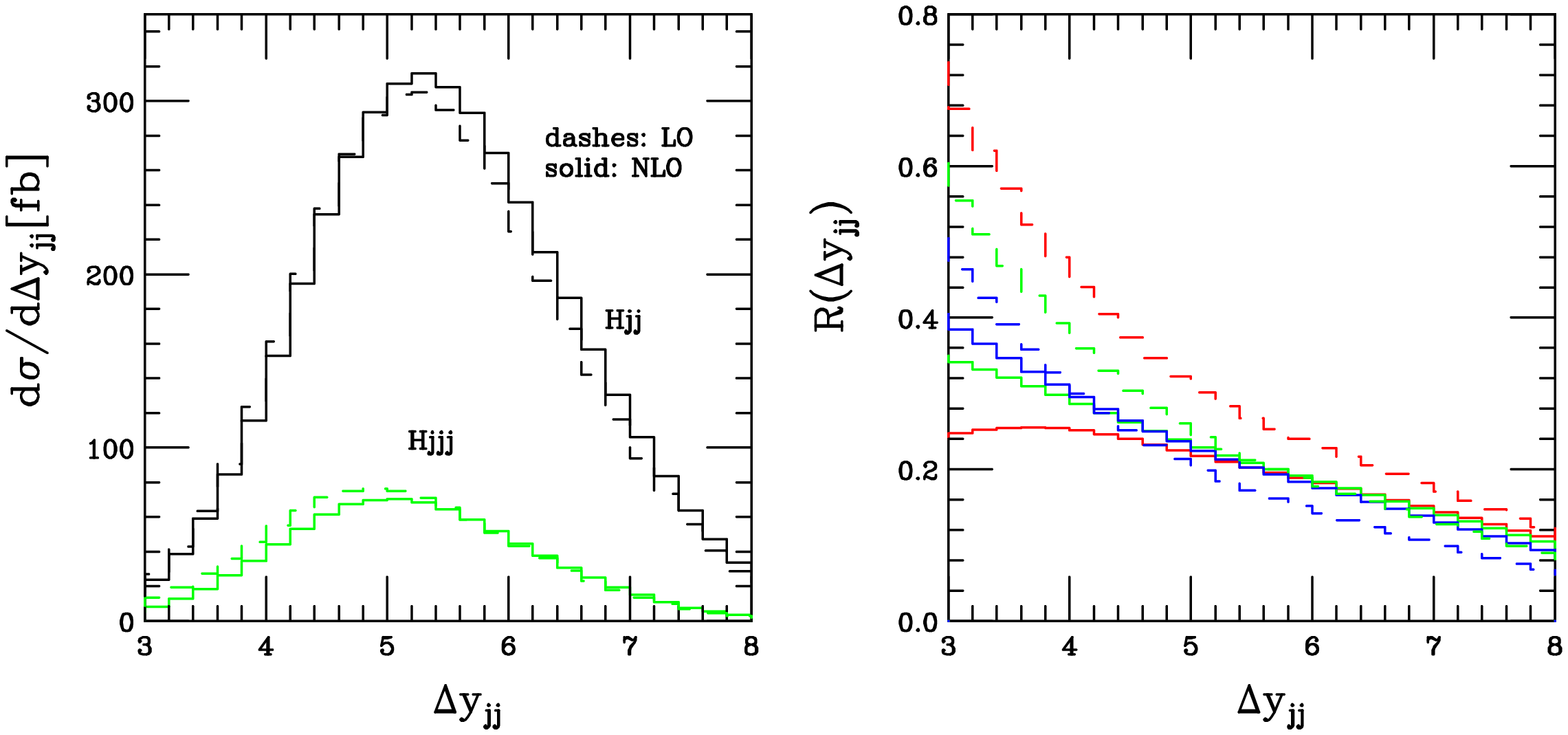,angle=90,width=5.5in}
\caption{The rapidity separation of two tagging jets for $m_h=120~\rm{GeV}$ 
within the cuts of Eqs.~(\ref{cuts1})-(\ref{eq:ophem}) and
Eq.~(\ref{eq:dijet}).  
In the left panel, $d\sigma/d\Delta y_{jj}$ is shown at 
NLO (solid histograms) and LO (dashed histograms) for $Hjj$ and $Hjjj$ 
production with a fixed scale  $\mu_R=\mu_F=40~{\rm GeV}$. 
In the right panel, $3$-jet ratios, $R(\Delta y_{jj})$ are shown at LO
(dashed) and at NLO (solid) for $\mu_R=\mu_F=20,40$, and $80~{\rm GeV}$. }%
 \label{dsdyjj}} 

When contemplating a central jet veto for the VBF signal, the probability  
for observing three (or more) jets in the final state becomes crucial. 
With the two leading jets defined as tagging jets, one would like to
know this probability for emitting additional jets as a function of 
tagging jet distributions. It is given by the $3$-jet ratio 
$R=\sigma_{3}/\sigma_{2}$, which we define for arbitrary 
distributions as
\beq
R^{\{LO,NLO\}}(x) = \frac{d \sigma_{3}^{\{LO,NLO\}}(\mu_{R},\mu_{F})/dx}
                   {d \sigma_{2}^{NLO}(\mu_{R}=\mu_{F}=m_{h})/dx}.
\eeq
For both NLO and LO $3$-jet ratios, the distribution for Higgs plus two 
jet production in the denominator
is computed to NLO accuracy since this provides the most accurate cross 
section estimate. For these $Hjj$ distributions, the NLO parton-level 
Monte Carlo program described in \cite{Figy:2003nv} is used with 
renormalization scale and factorization scale set to the mass of the Higgs 
boson, $m_{h}$. The numerator corresponds to the analogous distribution 
for Higgs plus three jet inclusive events (VBF $Hjjj$ production) for which 
we explore LO and NLO predictions and different scale choices. 

\FIGURE{
\epsfig{file=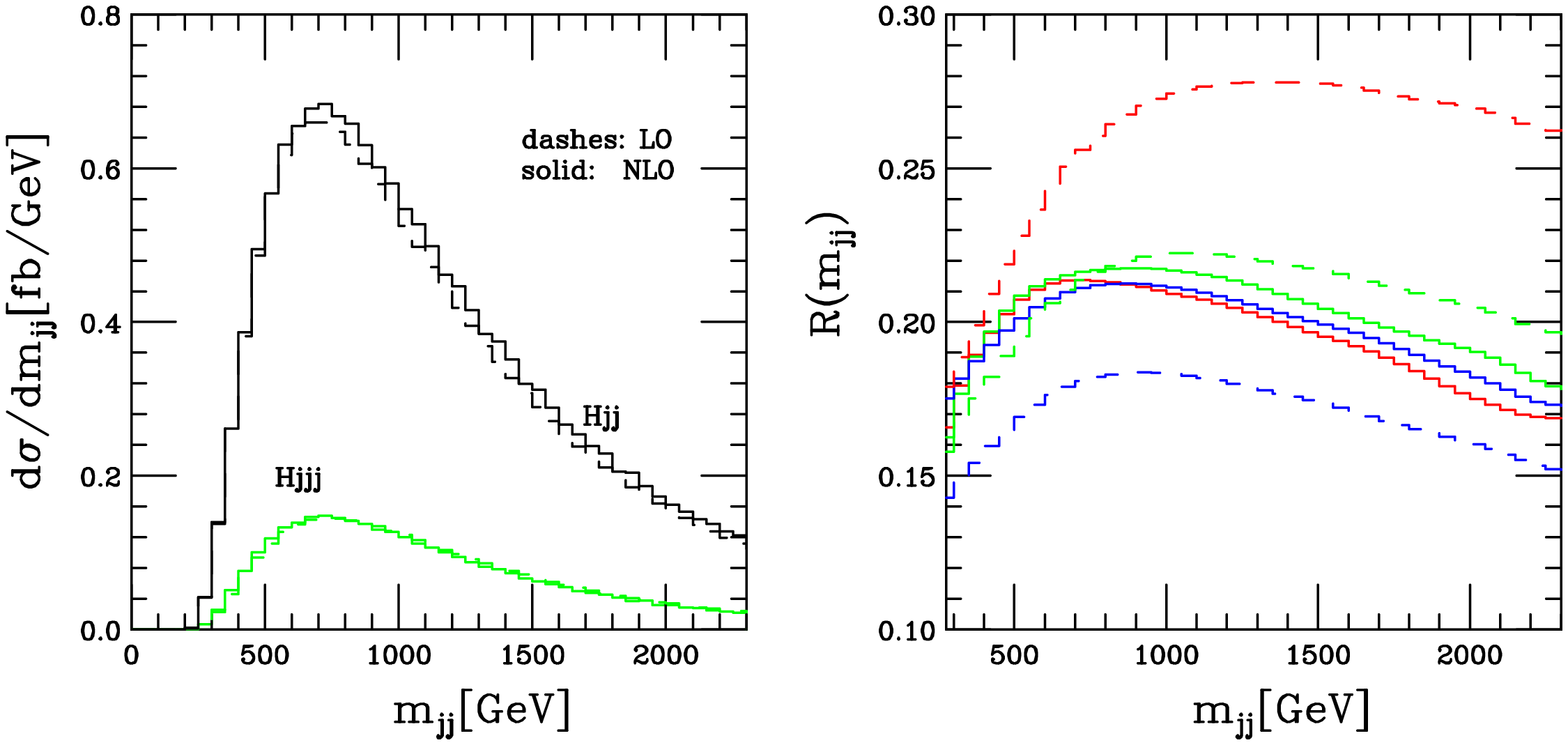,angle=90,width=5.5in}
\caption{The dijet invariant mass of two tagging jets for $m_h=120~\rm{GeV}$ 
within the cuts of Eqs.~(\ref{cuts1})-(\ref{eq:yjjcut}).
The meaning of the curves is the same as in Fig.~\ref{dsdyjj}.}
\label{dsdmjj}}

Let us start by considering the scale variations of the $3$-jet ratio 
as a function of the rapidity separation of the tagging jets, 
$x=\Delta y_{jj}$, (in Fig.~\ref{dsdyjj}) and of the invariant mass, 
$x=m_{jj}$, of the two tagging jets (in Fig.~\ref{dsdmjj}). The left-hand
panels show the distributions for 2-jet inclusive and 3-jet inclusive events
as predicted at LO (dashed histograms) and NLO (solid histograms) for a fixed
scale $\mu_R=\mu_F=40~{\rm GeV}$. The right-hand panels then give the 
corresponding $3$-jet ratios for three choices of scales, 
$\mu_{R} =\mu_{F} = 20,40$, and $80 ~{\rm GeV}$.  

The 3-jet ratio decreases with increasing rapidity separation of the tagging
jets. This is largely a kinematic effect: additional radiation in VBF events 
is mostly emitted outside the rapidity range set by the two tagging jets. 
Thus, the available phase space for additional jets diminishes rapidly
as $\Delta y_{jj}$ increases.
While typical 3-jet ratios are around 0.2, the LO ratio 
$R^{LO}(\Delta y_{jj})$ (dashed curves) reaches values up to $0.7$ at 
low values of $\Delta y_{jj}$. The corresponding NLO ratio is significantly
lower, around $0.4$. The reason is that at NLO the separation of the tagging
jets increases somewhat. When normalizing to the NLO $Hjj$ cross section
$R^{LO}$ is enhanced in the $\Delta y_{jj}=3$ region, where cross sections 
are very small due to the $m_{jj}>600$~GeV cut. There is no such effect 
for $R^{NLO}$. One also notices that for higher values of $\Delta y_{jj}$ 
the scale dependence decreases, becoming insignificant at NLO in the 
phase space region with typical VBF cuts ($\Delta y_{jj}>4$).

Similar threshold effects appear in the $m_{jj}$ distributions of 
Fig.~\ref{dsdmjj}: large scale variations at NLO are confined to 
the low $m_{jj}$ region with negligible cross section due to the cuts.
The 3-jet ratios decrease somewhat at large values of the dijet invariant
mass. However, the effect is not as strong as in the $\Delta y_{jj}$ 
distribution. Particularly striking is the reduction of the scale uncertainty
when going from $R^{LO}$ ($\approx 30\%$) to $R^{NLO}$ (5 to 10\%).

Veto jets are typically defined to be non-tagging jets that reside in 
the rapidity region between the tagging jets.  In addition to the cuts 
of  Eqs.~(\ref{cuts1})-(\ref{eq:dijet}), we employ the following 
definition for the veto jets,
\beq \label{eq:veto}
p_{Tj}^{\rm veto} > p_{T,veto}\, , \quad \quad 
y_{j}^{\rm veto} \in (y_{j}^{\rm tag~1},y_{j}^{\rm tag~2}) \, .
\eeq
For 4-jet events it is possible to identify two veto jets. In this case, 
we order the veto jets according to their 
transverse momentum with $p_{Tj}^{\rm veto~1} > p_{Tj}^{\rm veto~2}$. 
In the following we take $p_{T,veto}=20~{\rm GeV}$ unless stated otherwise.

On the left-hand-side of Fig.~\ref{dsdyrel} the rapidity distribution, 
$d\sigma/dy_{rel}$, of the highest $p_{T}$ veto jet is shown. Here the 
rapidity is measured with respect to the average rapidity of the tagging 
jets, 
\beq\label{eq:defyrel}
y_{rel}=y_{j}^{\rm veto}-(y_{j}^{\rm tag~1}+y_{j}^{\rm tag~2})/2 \,.
\eeq
The two histograms correspond to the LO (dashed green) and NLO (solid red) 
distributions at a scale $\mu_R=\mu_F=40~{\rm GeV}$. The suppression of jet 
activity in the center, near $y_{rel}=0$, is even more pronounced at NLO than 
at LO, i.e. the higher order corrections strengthen the rapidity gap 
features of VBF events. This is reflected by the $K$ factor (solid line in 
right-side panel of Fig.~\ref{dsdyrel}) which is greater than one for 
$|y_{\rm rel}|>2$ and is less than one in the central region between 
the tagging jets. The right-hand-side of the figure also shows the 
scale variations for $\xi = 2^{\pm 1}$ relative 
to the $\xi=1$ case: the scale dependence is significantly reduced at NLO 
(dashed curves) and remains largest in the regions of small cross section.
In the vicinity of $y_{\rm rel}=0$ the NLO result varies between $-20\%$ 
and $+7\%$ down from a LO variation of $-20\%$ to $+24\%$. In the large 
cross section regions, near $y_{\rm rel}\approx \pm 2$, the scale 
variations at NLO are a few percent only, a drastic improvement from 
the LO situation. This small scale dependence in the large cross section 
region will be reflected in small QCD uncertainties at NLO for jet veto 
probabilities.

The effect is clearly visible in Fig.~\ref{dsdptveto} where the transverse 
momentum distribution for the highest $p_{T}$ veto jet is shown for 
$\mu_R=\mu_F=40~{\rm GeV}$ at LO (dashed green) and NLO (solid red).
The scale variations are largest at high $p_{Tj}^{\rm veto}$, but even at 
a value of $80~{\rm GeV}$ the NLO results for $\xi = 2^{\pm 1}$ 
(dashed curves) deviate from the $\xi=1$ case by only $-3\%$ to $+10\%$.
At LO (dotted) these scale variations are $-22\%$ to $+31\%$. 
The $K$ factor (solid green) is close to one but decreases monotonically,
i.e. at NLO the veto jet becomes slightly softer.   

\FIGURE{
\epsfig{file=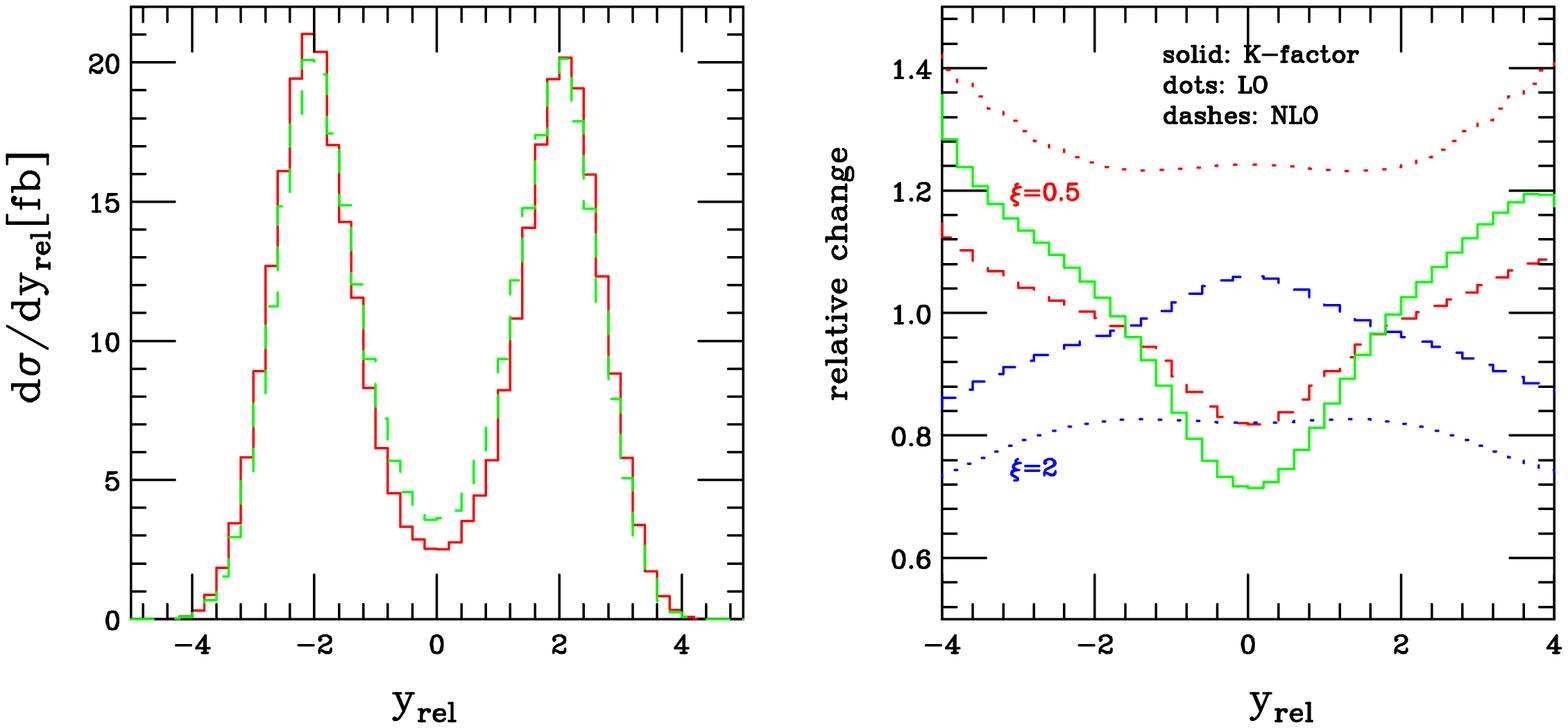,angle=90,width=5.5in}
\caption{The distribution in rapidity of the highest $p_{T}$ veto jet 
with the cuts of Eqs.~(\ref{cuts1})-(\ref{eq:dijet}) and Eq.~(\ref{eq:veto}), 
measured with respect to the rapidity average of the tagging jets.
In the left panel, $d\sigma/dy_{rel}$ is shown at LO (dashed green) 
and NLO (solid red) for $\mu_F = \mu_R = 40~{\rm GeV}$.  In the right-hand 
panel the $K$ factor (solid green) and scale variations of LO (dotted) and 
NLO (dashed) results are shown for $\mu_R = \mu_F=\xi \mu_{0}$ with 
$\xi = 1/2$ and $2$. }%
\label{dsdyrel}}

\FIGURE{
\epsfig{file=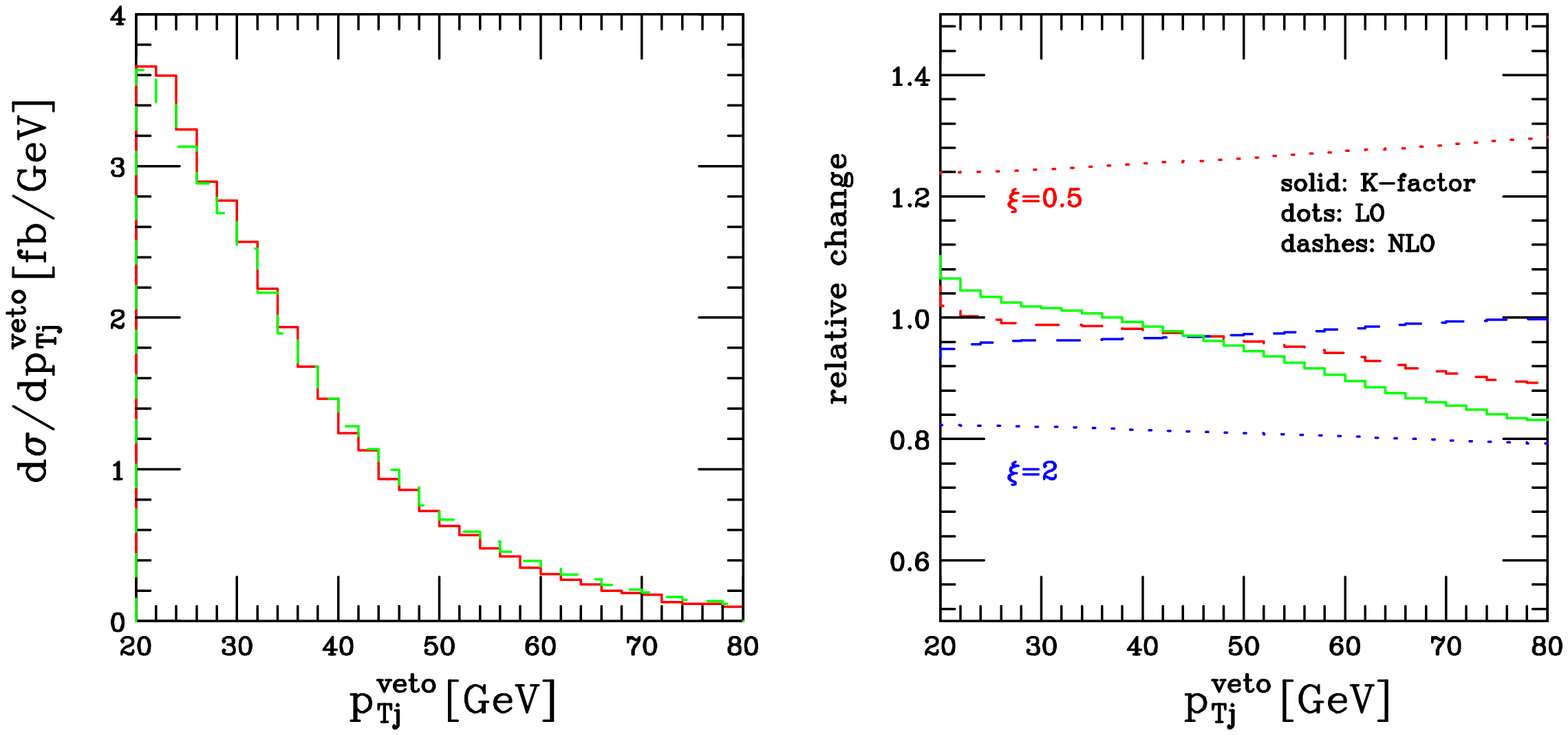,angle=90,width=5.5in}
\caption{Same as Fig.~\ref{dsdyrel} but for the distribution in 
the transverse momentum, $p_{Tj}^{\rm veto}$, 
of the highest $p_{T}$ veto jet.
}%
\label{dsdptveto}}

\FIGURE{
\epsfig{file=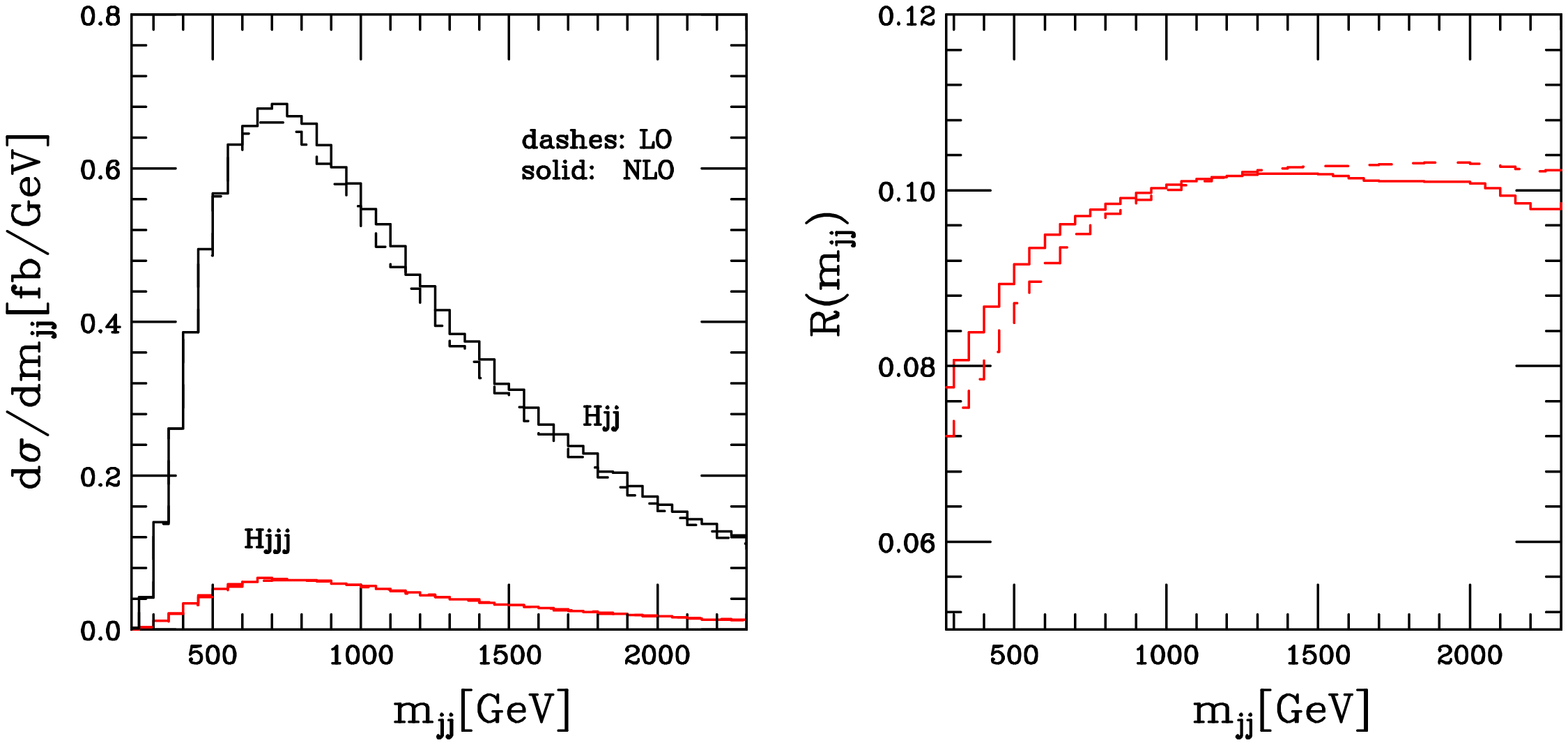,angle=90,width=5.5in}
\caption{The invariant mass distribution of the two tagging jets for 
$m_h=120~\rm{GeV}$ within the cuts of 
Eqs.~(\ref{cuts1})-(\ref{eq:ophem}), Eq.~(\ref{eq:yjjcut}) and 
Eq.~(\ref{eq:veto}).  
In the left panel, $d\sigma/d m_{jj}$ is shown at NLO (solid) and LO 
(dashed) for $Hjj$ and for $Hjjj$ production at $\mu_R=\mu_F=40 ~{\rm GeV}$. 
In the right panel, the corresponding $3$-jet ratios, $R^{LO}(m_{jj})$ 
(dashed) and $R^ {NLO}( m_{jj})$ (solid) are shown for the same scale 
choice. }%
\label{mjjcjet}}

\FIGURE{
\epsfig{file=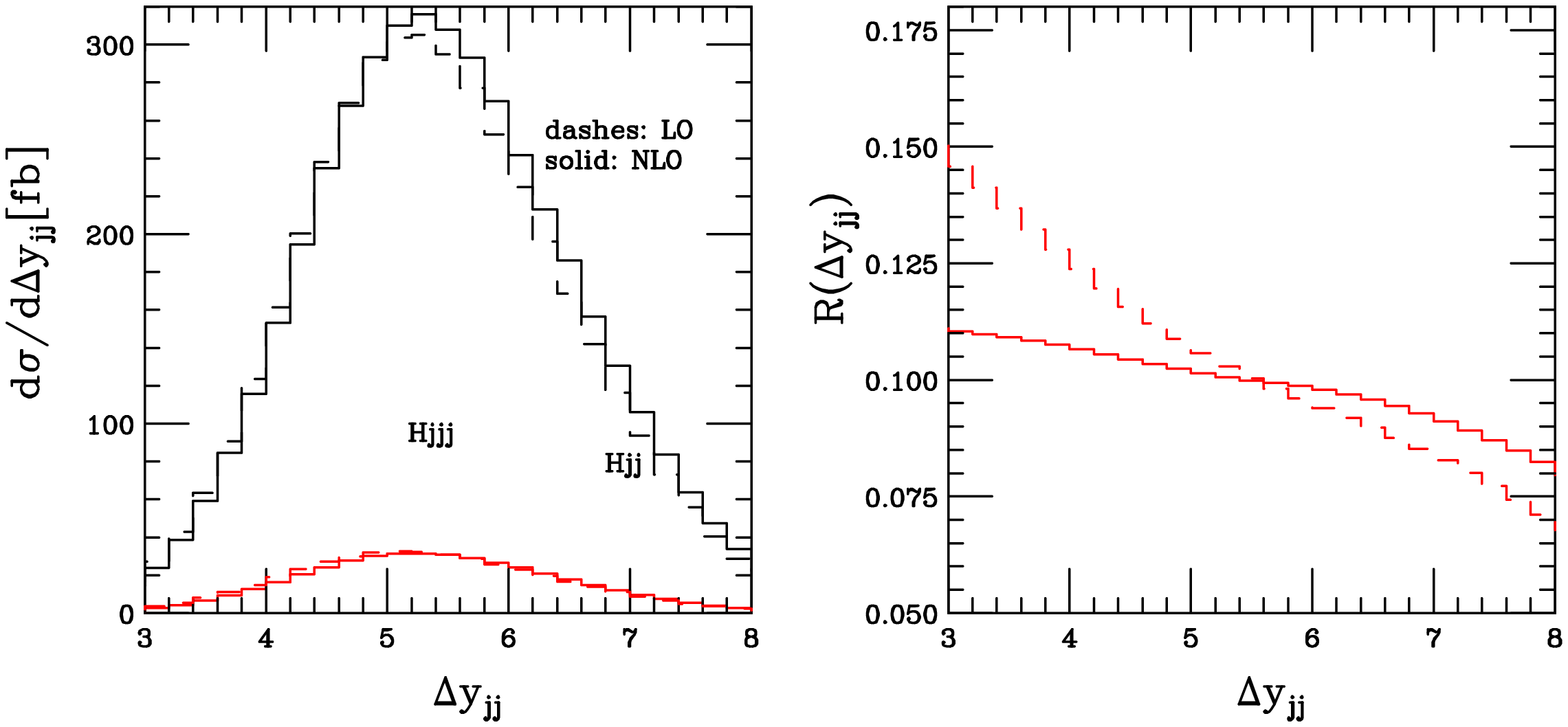,angle=90,width=5.5in}
\caption{Same as Fig.~\ref{mjjcjet} but for the rapidity separation of 
the two tagging jets and within the cuts of 
Eqs.~(\ref{cuts1})-(\ref{eq:ophem}), Eq.~(\ref{eq:dijet}), 
and Eq.~(\ref{eq:veto}). 
}%
\label{yjjcjet}}

\FIGURE{
\epsfig{file=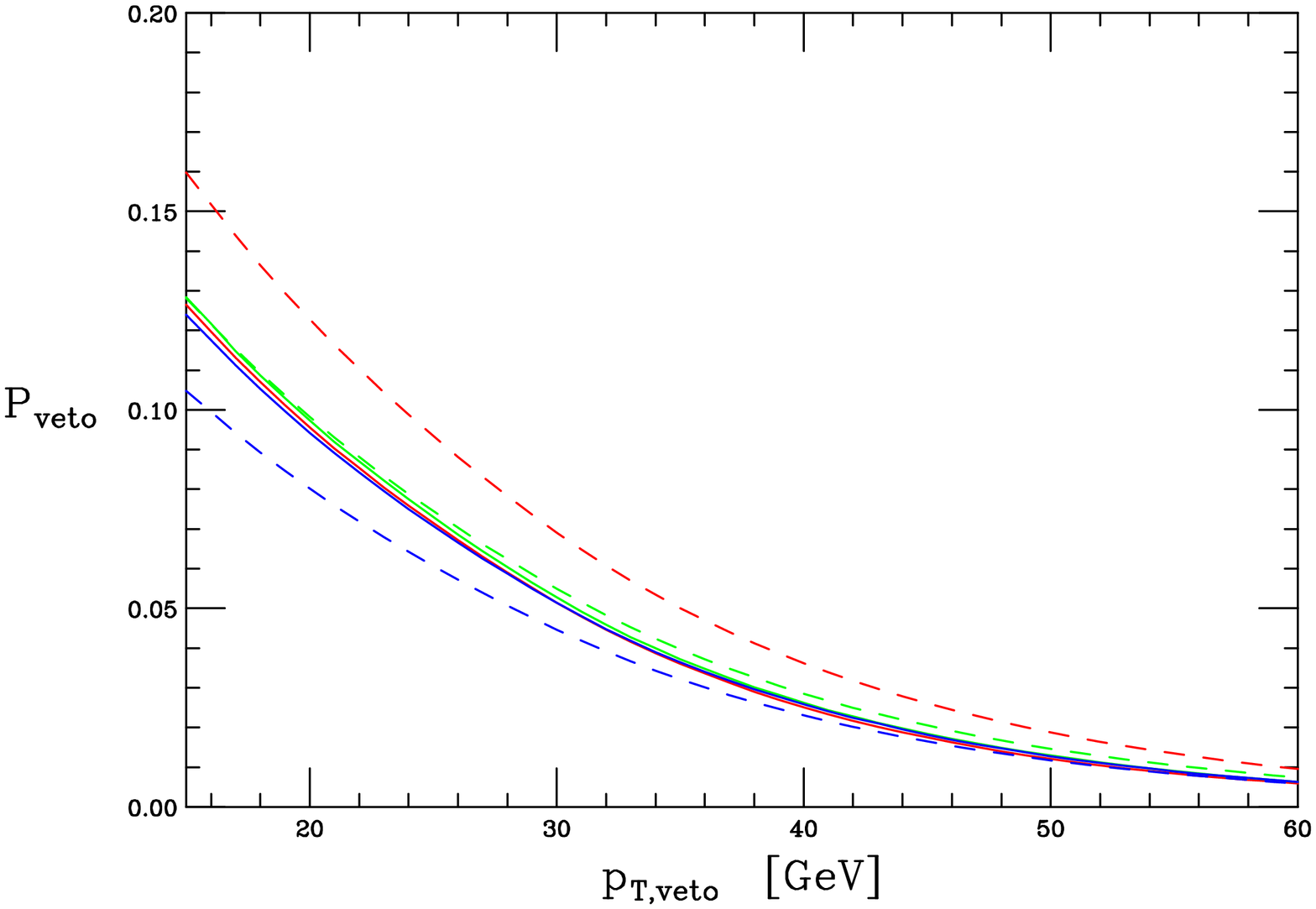,angle=90,width=4.0in}
\caption{ Ratio of the 3-jet cross section to the NLO 2-jet cross section 
for VBF, $P_{\rm veto} = \sigma_{3}/\sigma^{NLO}_{2}$. The dashed curves 
depict LO ratios and solid curves depict NLO ratios for the following 
scale choices: $\mu_{R}=\mu_{F}=20~{\rm GeV}$ (red),
$\mu_{R}=\mu_{F}=40~{\rm GeV}$ (green), and  
$\mu_{R}=\mu_{F}=80~{\rm GeV}$ (blue).} %
\label{nveto}}

Fig.~\ref{mjjcjet} shows the effect of the veto cuts defined by 
Eq.~(\ref{eq:veto}) on the tagging jet invariant mass distribution. 
Both LO and NLO 3-jet ratios are reduced compared to Fig.~\ref{dsdmjj}
due to the restricted rapidity range of Eq.~(\ref{eq:veto}) for the 
veto jets. Fig.~\ref{yjjcjet} depicts 
the distribution in rapidity separation of the tagging jets with veto 
cuts. Again, the 3-jet ratios are reduced. However, one also finds a 
significant shape change of the $\Delta y_{jj}$ dependence: 
the fairly steep decrease of the 3-jet ratio with increasing $\Delta y_{jj}$
becomes much less pronounced at NLO.

In Fig.~\ref{nveto} we show the probability for finding a veto jet,
\beq \label{nratio}
P_{\rm veto} = P(p_{T,veto}) = \frac{1}{\sigma_{2}^{NLO}} 
\int_{p_{T,veto}}^{\infty} dp_{Tj}^{veto} 
\frac{d \sigma_{3}}{d p_{Tj}^{veto}}
\eeq
as a function of the minimum transverse momentum of the hardest veto jet, 
$p_{T,veto}$. 
The scale variations at LO for the absolute veto probability are on the 
order of up to $\pm 3\%$. The NLO corrections reduce this scale dependence 
to below the $1\%$ level, i.e. to a negligible uncertainty.
When imposing a central jet veto, the accepted VBF Higgs production cross
section is given by
\beqn
\sigma_{2}({\rm veto}) = (1 - P_{\rm veto}) \sigma_{2}.
\eeqn 
Since, at NLO, $P_{\rm veto}$ is only about 10\% for a veto jet $p_T$ 
threshold of 20~GeV (and lower for harder thresholds) the perturbative
uncertainty on the SM prediction for the Higgs cross section due to 
a central jet veto is of order 1\% only at NLO and hence negligible 
compared to expected statistical errors~\cite{Zeppenfeld:2000td}.

%% file: conclusions.tex
\section{Discussion and Conclusions}
\label{sec:concl}
In this paper we have presented the dominant QCD corrections for 
Higgs production via vector-boson fusion in association with three jets.
The calculations are implemented in the form of a parton-level 
Monte Carlo program which allows to analyze arbitrary collinear and
infrared safe distributions with NLO QCD accuracy. 

Our calculation involves several approximations which significantly
reduce the complexity of the virtual corrections. Since we are only 
interested in phase space regions where vector boson fusion processes
can be distinguished from QCD backgrounds, we neglect contributions 
which are small once typical VBF cuts, in particular wide tagging jet
separations and large tagging jet invariant masses, are imposed.
Identical fermion interference effects are small after VBF cuts and 
we have also estimated the contribution from $t$-channel gluon 
exchange in virtual diagrams (and related real emission diagrams) 
to be well below one permille over the entire phase space relevant for
VBF production. Neglecting 
these small contributions, the QCD 1-loop corrections involve only 
a single quark line and are similar in complexity to dijet production 
in DIS~\cite{DIS2jNLO}, i.e. they
require the calculation of box diagrams as the most complex ingredient.

One reason for the smallness of the $t$-channel gluon contributions 
is that they are color suppressed, by a factor of $1/(N^2-1)$ in an 
$SU(N)$ gauge theory, and this feature is generic since gluon colors
need to be correlated to match the color singlet exchange nature of the 
tree level VBF process. In addition, we find very strong 
kinematical suppression factors in our analysis of the $t$-channel
gluon contributions which can be traced to the characteristic gluon
radiation pattern in VBF events. It is this kinematical suppression 
which renders the $t$-channel gluon contributions truly negligible.
It would be interesting to find out, whether this kinematical
suppression persists at higher orders.
  
In our phenomenological analysis for the LHC we find that
additional jet activity between the tagging jets in VBF Higgs 
production events is even more strongly suppressed once NLO QCD
corrections are included: $K$ factors go down to 0.7 for jet emission 
at the center between the two tagging jets. 
This strengthens the case for a central
jet veto as a background suppression tool. Requiring the absence 
of any additional jet activity of $p_{Tj}>20$~GeV between the 
the tagging jets we find veto probabilities for the signal of 
$P_{\rm veto}\approx 10\%$ from this perturbative QCD 
source. Our NLO QCD predictions for the veto probability show small 
residual scale variations, indicating a relative error on 
$P_{\rm veto}$ due to higher order effects of 10\% or less. This 
implies that the survival probability $P_{\rm surv} = 1-P_{\rm veto}$
can be determined with a perturbative QCD uncertainty of 
about 1\%, which is more than sufficient for Higgs coupling
determinations at the LHC~\cite{Zeppenfeld:2000td}.

Beyond the additional jet activity from perturbative QCD radiation, 
which we have analyzed in this paper, additional central jets in VBF
events will arise from multiple parton scattering (i.e. the underlying 
event) and from pile-up in high luminosity running. For small veto
thresholds $p_{T,veto}$, the contributions 
from these sources may be as large as the perturbative effects which 
we have considered and need to be estimated independently. However,
these additional contributions should be independent of the hard 
scattering event and can, hence, be determined from other LHC data, in 
particular by measuring the jet activity in other VBF processes. What 
will be needed on the 
theoretical side, is a precise calculation of the perturbative
contribution to the veto probability for these other VBF processes,
similar to the calculation described in this paper.

%% file: appendix.tex
\section{Virtual Corrections}\label{app:A}
In this appendix, we give the expressions for the finite, reduced 
amplitudes, $\widetilde{\mc{M}}_{\tau}^{(i)}$
that appear in Eq.~(\ref{eq:boxline}) in terms of $\widetilde{B}_{0}$, 
$\widetilde{C}_{0}$, and $\widetilde{D}_{ij}$ functions.
Here $\widetilde{B}_{0}$, $\widetilde{C}_{0}$, and $\widetilde{D}_{ij}$ 
denote the finite parts of the Passarino-Veltman 
$B_{0}$, $C_{0}$, and $D_{ij}$ functions \cite{Passarino:1978jh}, and 
are given explicity below. We write
\beqn
\widetilde{\mathcal{M}}_{\tau}^{(1)} (k_2,q_1,q_2;\eps_{1},\eps_{2})&=& 
\bar{\psi}(k_2) \{ c_{q}^{(1)} (\slashit{q}_{1}-\slashit{q}_{2}) +  
c_{1}^{(1)} \slashit{\eps}_1 + c_{2}^{(1)} \slashit{\eps}_2 \nonumber \\  
&+& c_{b}^{(1)} \slashit{\eps}_2 (\slashit{k}_2 + \slashit{q}_2)
\slashit{\eps}_1 \} P_{\tau} \psi(k_1) \,, 
\eeqn
\beqn
\widetilde{\mathcal{M}}_{\tau}^{(2)} (k_2,q_1,q_2;\eps_{1},\eps_{2})&=&
\bar{\psi}(k_2) \{ c_{q}^{(2)} (\slashit{q}_{1}-\slashit{q}_{2}) +
c_{1}^{(2)} \slashit{\eps}_1 + c_{2}^{(2)} \slashit{\eps}_2 \nonumber \\
&+& c_{b}^{(2)} \slashit{\eps}_1 (\slashit{k}_2 + \slashit{q}_1)
\slashit{\eps}_2 \} P_{\tau} \psi(k_1)\, , 
\eeqn
and,
\beqn
 \widetilde{\mathcal{M}}_{\tau}^{(3)} (k_2,q_1,q_2;\eps_{1},\eps_{2})&=&
 \bar{\psi}(k_2) \{ c_{q}^{(3)} (\slashit{q}_{1}-\slashit{q}_{2}) +
c_{1}^{(3)} \slashit{\eps}_1 + c_{2}^{(3)} \slashit{\eps}_2 \nonumber \\
&+&c_{b}^{(3)}(u) \slashit{\eps}_1 (\slashit{k}_2 + \slashit{q}_1)
\slashit{\eps}_2 
+ c_{b}^{(3)}(t) \slashit{\eps}_2 (\slashit{k}_2 + \slashit{q}_2)
\slashit{\eps}_1 \} P_{\tau} \psi(k_1) \, ,
\eeqn
where $\eps_{1}=\eps_{1}(q_{1})$ and $\eps_{2}=\eps_{2}(q_{2})$ are the 
effective polarization vectors for the 
gluon and weak boson. The coefficient functions, $c_{i}^{(j)}$, 
with $j=1,2,3$ and $i=b,q,1,2$ are given below.
\beqn
c_{b}^{(1)}&=& \mathbf{Box}_{b}^{(1)} - \frac{2~\widetilde{B}_{0}(t)}{t} - 
\frac{T_{b}(q_{2}^{2},t)}{t}\\
c_{1}^{(1)}&=& \mathbf{Box}_{1}^{(1)} + 2~ \epsbkb T_{\epsilon}(\qbs,t) 
-2~ \epsbqb 
\frac{[\widetilde{B}_{0}(t)-\widetilde{B}_{0}(\qbs)]}{t-\qbs}\\
c_{2}^{(1)} &=& \mathbf{Box}_{2}^{(1)} + 2~ \epsaka T_{\epsilon}(0,t)\\
c_{q}^{(1)}&=& \mathbf{Box}_{q}^{(1)}
\eeqn
\beqn
c_{b}^{(2)} &=& \mathbf{Box}_{b}^{(2)} - \frac{2~\widetilde{B}_{0}(u)}{u} 
- \frac{T_{b}(q_{2}^{2},u)}{u}\\
c_{1}^{(2)} &=& \mathbf{Box}_{1}^{(2)} + 2~ \epsbka T_{\epsilon}(\qbs,u) 
+2~ \epsbqb 
\frac{[\widetilde{B}_{0}(u)-\widetilde{B}_{0}(\qbs)]}{u-\qbs}\\
c_{2}^{(2)} &=& \mathbf{Box}_{2}^{(2)} + 2~ \epsakb T_{\epsilon}(0,u)\\
c_{q}^{(2)}&=& \mathbf{Box}_{q}^{(2)}
\eeqn
\beqn
t c_{b}^{(3)}(t) &=& t \mathbf{Box}_{b}^{(3)} -2(t \widetilde{C}_{0}(t)+1) 
+ \widetilde{B}_{0}(t)+ T_{b}(\qbs,t)\\
u c_{b}^{(3)}(u) &=& u \mathbf{Box}_{b}^{(3)} -2( u \widetilde{C}_{0}(u)+1) 
+ \widetilde{B}_{0}(u)+T_{b}(\qbs,u)\\
c_{1}^{(3)} &=& \mathbf{Box}_{1}^{(3)} -2 \epsbkb T_{\epsilon}(\qbs,t) 
+ 2 \epsbqb 
\frac{[\widetilde{B}_{0}(t)-\widetilde{B}_{0}(\qbs)]}{t-\qbs}\\ \nonumber 
&-&2 \epsbka T_{\epsilon}(\qbs,u)
-2 \epsbqb \frac{[\widetilde{B}_{0}(u)-\widetilde{B}_{0}(\qbs)]}{u-\qbs}\\
c_{2}^{(3)} &=& \mathbf{Box}_{2}^{(3)} 
+ \frac{2}{t} (t\widetilde{C}_{0}(t)+1) \epsaka +
\frac{2}{u} (u \widetilde{C}_{0}(u)+1) \epsakb \\
c_{q}^{(3)}&=&\mathbf{Box}_{q}^{(3)}
\eeqn
The $T_i$ functions are explicitly listed below.
\beqn
T_{b}(q^2,t) &= &\frac{1}{t-q^2} \{2 q^2 [\widetilde{B}_{0}(t) - 
\widetilde{B}_{0}(q^2)] + t \widetilde{B}_{0}(t) \nonumber \\
&&-q^2 \widetilde{B}_{0}(q^2)\} - 2q^2 \widetilde{C}_{0}(q^2,t) \\
T_{\epsilon}(q^2,t)& = & \frac{1}{t-q^2} \big \{ [\widetilde{B}_{0}(t) - 
\widetilde{B}_{0}(q^2)]\frac{2t + 3 q^2}{t-q^2} +
2 \widetilde{B}_{0}(q^2) + 1 - 2q^2 \widetilde{C}_{0}(q^2,t) \big \}\\
T_{\epsilon}(0,t)&=& \frac{1}{t}(2 \widetilde{B}_{0}(t)+1)\\
T_{b}(0,t)&=& \widetilde{B}_{0}(t)
\eeqn
Here the coefficients, $\mathbf{Box}_{i}^{j}$, for $j=1,2,3$ and $i=b,q,1,2$
are expressed in terms of the Passarino-Veltman $\td{ij}$ functions. 
The $\mathbf{Box}$ coefficients with
$\td{ij} = \td{ij}(q_1,k_2,q_2)$ are listed below.
\begin{eqnarray}
\mathbf{Box}_{b}^{(3)} &=& (6\td{27}+\frac{3}{2}\td{0}\qbs+\frac{5}{2}\td{12}\qbs - \td{13}\qbs-3\td{23}\qbs+\td{24}\qbs - 2\td{25}\qbs \nonumber \\
&+&4\td{26}\qbs+\frac{1}{2}\td{0}t +3\td{11}t-\frac{5}{2}\td{12}t+\td{21}t -\td{24} t + 4\td{25}t \nonumber\\
&-& 4\td{26}t-\frac{3}{2}\td{0}u -\td{11}u-\frac{9}{2}\td{12}u+4\td{13}u +\td{21}u \nonumber \\
&-&5\td{24}u+4\td{25}u)/2 
\end{eqnarray}

\begin{eqnarray}
\mathbf{Box}_{q}^{(3)} &=& -\td{27} \epsaepsb -2 \td{311} \epsaepsb + 2 \td{313} \epsaepsb - 8 \td{12} \epsakb \epsbkb \nonumber \\
&+& 8 \td{13} \epsakb \epsbkb - 4 \td{22} \epsakb \epsbkb -8 \td{24} \epsakb \epsbkb + 12 \td{26} \epsakb \epsbkb \nonumber \\
&-& 4 \td{36} \epsakb \epsbkb + 4 \td{38} \epsakb \epsbkb + \frac{3}{2} \td{0} \epsaqb \epsbkb + \frac{3}{2} \td{12} \epsaqb \epsbkb \nonumber \\
&+& 4 \td{23} \epsaqb \epsbkb - 8 \td{25} \epsaqb \epsbkb + 4 \td{26} \epsaqb \epsakb - 4 \td{310} \epsaqb \epsbkb \nonumber \\
&+& 4 \td{39} \epsaqb \epsbkb-\frac{3}{2} \td{0} \epsakb \epsbqa -\frac{19}{2} \td{12} \epsakb \epsbqa + 8 \td{13} \epsakb \epsbqa \nonumber \\
&-& 12 \td{24} \epsakb \epsbqa + 8 \td{25} \epsakb \epsbqa + 4 \td{26} \epsakb \epsbqa+4 \td{310} \epsakb \epsbqa \nonumber \\
&-& 4 \td{34} \epsakb \epsbqa + 4 \td{23} \epsaqb \epsbqa-4 \td{25} \epsaqb \epsbqa -4 \td{35} \epsaqb \epsbqa \nonumber \\
&+& 4 \td{37} \epsaqb \epsbqa-\frac{3}{2} \td{0} \epsakb \epsbqb -\frac{11}{2} \td{12} \epsakb \epsbqb + 4 \td{13} \epsakb \epsbqb \nonumber \\
&+& 8 \td{23} \epsakb \epsbqb - 4 \td{24} \epsakb \epsbqb - 4 \td{26} \epsakb \epsbqb-4 \td{310} \epsakb \epsbqb \nonumber \\
&+& 4 \td{39} \epsakb \epsbqb + 4 \td{23} \epsaqb \epsbqb-4 \td{25} \epsaqb \epsbqb + 4 \td{33} \epsaqb \epsbqb \nonumber \\
&-& 4 \td{37} \epsaqb \epsbqb - \frac{1}{2}\td{12} \epsaepsb \qbs + \frac{1}{2}\td{13} \epsaepsb \qbs + \frac{1}{2}\td{23} \epsaepsb \qbs)\nonumber \\
&-& \frac{1}{2}\td{24} \epsaepsb \qbs -\td{310} \epsaepsb \qbs -\td{33} \epsaepsb \qbs+\td{37} \epsaepsb \qbs \nonumber \\
&+&\td{39} \epsaepsb \qbs-\td{0} \epsaepsb t -\frac{3}{2}\td{11} \epsaepsb t +\frac{1}{2}\td{12} \epsaepsb t \nonumber \\
&-& \frac{1}{2}\td{21} \epsaepsb t  +\frac{1}{2}\td{24} \epsaepsb t +\td{310} \epsaepsb t-\td{35} \epsaepsb t \nonumber \\
&+ &\td{37}\epsaepsb t - \td{39}\epsaepsb t  -\frac{11}{2}\td{11}\epsaepsb u +\frac{3}{2}\td{12}\epsaepsb u \nonumber \\ 
&-& \td{13}\epsaepsb u +\frac{1}{2}\td{21}\epsaepsb u+\td{23} \epsaepsb u +\frac{3}{2}\td{24}\epsaepsb u \nonumber \\
&-& 2\td{25}\epsaepsb u-\td{26}\epsaepsb u -\td{310}\epsaepsb u +\td{34} \epsaepsb u \nonumber \\
&-&\td{35} \epsaepsb u + \td{37} \epsaepsb u
\end{eqnarray}
\begin{eqnarray}
\mathbf{Box}_{1}^{(3)}& = &24\td{27} \epsbkb+20\td{312} \epsbkb+22\td{27}\epsbqa +20\td{311}\epsbqa \nonumber \\
&+& 12\td{27}\epsbqb+20\td{313}\epsbqb -4\td{23}\epsbkb \qbs+4\td{26}\epsbkb \qbs \nonumber \\
&+& 2\td{38}\epsbkb\qbs -2\td{39}\epsbkb\qbs+\td{12}\epsbqa\qbs-\td{13}\epsbqa\qbs \nonumber \\
&-&3\td{23}\epsbqa\qbs+\td{24}\epsbqa\qbs+2\td{26}\epsbqa\qbs +2\td{310}\epsbqa\qbs \nonumber \\
&-&2\td{37}\epsbqa\qbs-2\td{23}\epsbqb\qbs +2\td{26}\epsbqb\qbs-2\td{33}\epsbqb\qbs \nonumber \\
&+& 2\td{39}\epsbqb\qbs +\frac{3}{2}\td{0}\epsbkb t+\frac{3}{2}\td{12}\epsbkb t+4\td{13}\epsbkb t \nonumber \\
&+&2\td{25}\epsbkb t+2\td{26}\epsbkb t+2\td{310}\epsbkb t -2\td{38}\epsbkb t \nonumber \\
&+& 2\td{0}\epsbqa t+3\td{11}\epsbqa t -\td{12}\epsbqa t+4\td{13}\epsbqa t \nonumber \\
&+&\td{21}\epsbqa t -\td{24}\epsbqa t+6\td{25}\epsbqa t-2\td{26}\epsbqa t \nonumber \\
&-& 2\td{310}\epsbqa t+2\td{35}\epsbqa t+4\td{13}\epsbqb t +4\td{23}\epsbqb t \nonumber \\
&+&2\td{25}\epsbqb t-2\td{26}\epsbqb t +2\td{37}\epsbqb t-2\td{39}\epsbqb t \nonumber \\
&+&\frac{3}{2}\td{0}\epsbkb u + \frac{7}{2}\td{12}\epsbkb u -2\td{13}\epsbkb u+2 \td{22}\epsbkb u \nonumber \\
&-& 2\td{24}\epsbkb u+2\td{25}\epsbkb u-2\td{26}\epsbkb u +2\td{310}\epsbkb u \nonumber \\
&-&2\td{36}\epsbkb u+\frac{3}{2}\td{0}\epsbqa u +\td{11}\epsbqa u+\frac{5}{2}\td{12}\epsbqa u \nonumber \\
&-&2\td{13}\epsbqa u -\td{21}\epsbqa u+\td{24}\epsbqa u-2\td{34}\epsbqa u \nonumber \\
&+&2\td{35}\epsbqa u+\frac{3}{2}\td{0} \epsbqb u +\frac{3}{2}\td{12} \epsbqb u -2\td{23}\epsbqb u \nonumber \\
&+& 2\td{26}\epsbqb u-2\td{310}\epsbqb u + 2\td{37}\epsbqb u
\end{eqnarray}
\begin{eqnarray}
\mathbf{Box}_{2}^{(3)} &=& -12\td{27} \epsakb-4\td{312} \epsakb-6\td{27}\epsaqb -4\td{313}\epsaqb \nonumber \\
&-& 3\td{0}\epsakb\qbs-7\td{12}\epsakb\qbs +2\td{13}\epsakb\qbs-2\td{22} \epsakb\qbs \nonumber \\
&+& 6\td{23}\epsakb\qbs -2\td{24}\epsakb\qbs+4\td{25}\epsakb\qbs-8\td{26}\epsakb\qbs \nonumber \\
&-& 2\td{38}\epsakb\qbs+2\td{39}\epsakb\qbs-\frac{3}{2}\td{0}\epsaqb\qbs -\frac{5}{2}\td{12}\epsaqb\qbs \nonumber \\
&-& \td{13}\epsaqb\qbs+3\td{23}\epsaqb\qbs -\td{24}\epsaqb\qbs +2\td{25}\epsaqb\qbs \nonumber \\
&-& 6\td{26}\epsaqb\qbs +2\td{33}\epsaqb\qbs-2\td{39}\epsaqb\qbs +\frac{3}{2}\td{0}\epsakb t \nonumber \\
&-&4\td{11}\epsakb t+\frac{11}{2}\td{12}\epsakb t-2\td{21}\epsakb t +2\td{22}\epsakb t\nonumber \\
&-& 6\td{25}\epsakb t+6\td{26}\epsakb t -2\td{310}\epsakb t +2\td{38}\epsakb t \nonumber \\
&-& \frac{1}{2}\td{0}\epsaqb t -3\td{11}\epsaqb t+\frac{5}{2}\td{12}\epsaqb t -\td{21}\epsaqb t \nonumber \\
&+&\td{24}\epsaqb t-6\td{25}\epsaqb t+6\td{26}\epsaqb t -2\td{37}\epsaqb t \nonumber \\
&+& 2\td{39}\epsaqb t+\frac{7}{2}\td{0}\epsakb u  +\frac{23}{2}\td{12}\epsakb u -6\td{13}\epsakb u \nonumber \\
&-& 2\td{21}\epsakb u +2\td{22} \epsakb u +10\td{24}\epsakb u - 6\td{25}\epsakb u \nonumber \\
&-&2\td{26}\epsakb u -2\td{310}\epsakb u+2\td{36} \epsakb u + 2\td{0} \epsaqb u \nonumber \\
&-& \td{11} \epsaqb u+5\td{12} \epsaqb u -\td{21} \epsaqb u -2\td{23} \epsaqb u \nonumber \\
&+& 3\td{24}\epsaqb u + 2\td{26}\epsaqb u +2\td{310}\epsaqb u -2\td{37} \epsaqb u 
\end{eqnarray}
%
%
The $\mathbf{Box}$ coefficients with $\td{ij}=\td{ij}(k_2,q_2,q_1)$ are listed below.
\begin{eqnarray}
\mathbf{Box}_{1}^{(1)} &=& -\epsaqa [ -8 \td{27}-8 \td{312}- ( \td{11} -\td{12}+\td{13}-4 \td{22} +4 \td{24} ) \qbs+\td{11} s - \td{12} s\nonumber \\
&+&\td{13} s-4 \td{22} s+4 \td{24} s +\td{11} t-\td{12} t+\td{13} t -4 \td{22} t + 4 \td{24} t ]  \nonumber \\
&+&\epsbqa [ 8 \td{27} +8 \td{313}+ ( \td{11}-\td{12} +\td{13}-4 \td{22}+4 \td{24} ) \qbs -\td{11} t-3 \td{12} t \nonumber \\
&+& 3 \td{13} t - 4 \td{24} t+4 \td{26} t ] - \epsbkb [ 16 \td{311} -24 \td{312}-(\td{11}-\td{12} +\td{13}+4 \td{25} \nonumber \\
&-& 4 \td{26} -8 \td{310} - 4 \td{32}-4 \td{34} +4 \td{35}+8 \td{36}+4 \td{38}) \qbs + 5 \td{11} s-5 \td{12} s \nonumber \\
&+& \td{13} s + 4 \td{21} s-4 \td{24} s+4 \td{25} s -4 \td{26} s-8 \td{310} s+4 \td{35} s +4 \td{38} s -4 \td{12} t \nonumber \\
&+& 4 \td{13} t +4 \td{22} t-8 \td{24} t+8 \td{25} t -4 \td{26} t-4 \td{310} t - 4 \td{34} t \nonumber \\
&+& 4 \td{35} t+4 \td{36} t ]
\end{eqnarray}
\begin{eqnarray}
\mathbf{Box}_{2}^{(1)} &=&\epsakb [8 \td{311}-24 \td{313}-(\td{11} + 3 \td{12}-3 \td{13}-4 \td{22} +8 \td{24}-4 \td{25} + 4 \td{310} \nonumber \\
&-& 4 \td{37}-4 \td{38}+4 \td{39}) \qbs -\td{11} s+\td{12} s-5 \td{13} s - 8 \td{25} s + 4 \td{26} s -4 \td{37} s \nonumber \\
&+& 4 \td{39} s+4 \td{12} t-4 \td{13} t -4 \td{23} t +4 \td{24} t-4 \td{25} t + 4 \td{26} t + 4 \td{310} t-4 \td{37} t] \nonumber \\
&-& \epsaqb [ 8 \td{27}-8 \td{312}+24 \td{313} + (\td{11}+3 \td{12}-3 \td{13} -4 \td{22}+8 \td{24} - 4 \td{25} \nonumber \\
&+& 4 \td{310}-4 \td{37}-4 \td{38} +4 \td{39}) \qbs+\td{11} s-\td{12} s + 5 \td{13} s + 8 \td{25} s-4 \td{26} s \nonumber \\
&+&4 \td{37} s-4 \td{39} s-\td{11} t - 7 \td{12} t+7 \td{13} t + 4 \td{23} t - 8 \td{24} t+4 \td{25} t-4 \td{310} t \nonumber \\
&+&4 \td{37} t ] 
\end{eqnarray}
\begin{eqnarray}
\mathbf{Box}_q^{(1)} &=&
8 \td{312} \epsaepsb-8 \td{313} \epsaepsb+8 \td{12} \epsakb \epsbkb - 8 \td{13} \epsakb \epsbkb \nonumber \\ 
&+& 12 \td{24} \epsakb \epsbkb -12 \td{25} \epsakb \epsbkb + 4 \td{34} \epsakb \epsbkb-4 \td{35} \epsakb \epsbkb \nonumber \\
&+& \td{11} \epsaqb \epsbkb + 7 \td{12} \epsaqb \epsbkb-7 \td{13} \epsaqb \epsbkb+4 \td{22} \epsaqb \epsbkb \nonumber \\
&+& 8 \td{24} \epsaqb \epsbkb-4 \td{25} \epsaqb \epsbkb-8 \td{26} \epsaqb \epsbkb - 4 \td{310} \epsaqb \epsbkb \nonumber \\
&+& 4 \td{36} \epsaqb \epsbkb-\td{11} \epsakb \epsbqa +\td{12} \epsakb \epsbqa-\td{13} \epsakb \epsbqa \nonumber \\
&-& 4 \td{23} \epsakb \epsbqa - 4 \td{25} \epsakb \epsbqa+8 \td{26} \epsakb \epsbqa+4 \td{310} \epsakb \epsbqa \nonumber \\
&-& 4 \td{37} \epsakb \epsbqa-4 \td{23} \epsaqb \epsbqa+4 \td{26} \epsaqb \epsbqa + 4 \td{38} \epsaqb \epsbqa \nonumber \\
&-& 4 \td{39} \epsaqb \epsbqa-\td{11} \epsakb \epsaqa + 5 \td{12} \epsakb \epsaqa -5 \td{13} \epsakb \epsaqa \nonumber \\
&+& 8 \td{22} \epsakb \epsaqa -4 \td{25} \epsakb \epsaqa -4 \td{26} \epsakb \epsaqa -4 \td{310} \epsakb \epsaqa \nonumber \\
&+& 4 \td{36} \epsakb \epsaqa +4 \td{12} \epsaqb \epsaqa -4 \td{13} \epsaqb \epsaqa +8 \td{22} \epsaqb \epsaqa \nonumber \\
&-& 8 \td{26} \epsaqb \epsaqa +4 \td{32} \epsaqb \epsaqa - 4 \td{38} \epsaqb \epsaqa -4 \td{310} \epsaepsb \qbs \nonumber \\
&-& 2 \td{32} \epsaepsb \qbs + 2 \td{36} \epsaepsb \qbs+2 \td{37} \epsaepsb \qbs+4 \td{38} \epsaepsb \qbs \nonumber \\
&-& 2 \td{39} \epsaepsb \qbs-\frac{1}{2}\td{11} \epsaepsb s +\frac{1}{2}\td{12} \epsaepsb s - \frac{1}{2}\td{13} \epsaepsb s \nonumber \\
&-& 2 \td{25} \epsaepsb s+2 \td{26} \epsaepsb s + 2 \td{310} \epsaepsb s-2 \td{37} \epsaepsb s \nonumber \\
&-& 2 \td{38} \epsaepsb s + 2 \td{39} \epsaepsb s -2 \td{22} \epsaepsb t-2 \td{23} \epsaepsb t \nonumber \\
&+& 4 \td{26} \epsaepsb t+4 \td{310} \epsaepsb t -2 \td{36} \epsaepsb t - 2 \td{37} \epsaepsb t
\end{eqnarray}
\begin{eqnarray}
\mathbf{Box}_b^{(1)} 
&=& -4 \td{27}-12 \td{312}+12 \td{313} +4 \td{310} \qbs+2 \td{32} \qbs-2 \td{36} \qbs -2 \td{37} \qbs -4 \td{38} \qbs+2 \td{39} \qbs \nonumber \\
&-& 2 \td{0} s-\td{11} s-\td{12} s +\td{13} s+2 \td{25} s-2 \td{26} s -2 \td{310} s+2 \td{37} s+2 \td{38} s \nonumber \\
&-& 2 \td{39} s+2 \td{22} t+2 \td{23} t - 4 \td{26} t -4 \td{310} t+2 \td{36} t +2 \td{37} t
\end{eqnarray}
The $\mathbf{Box}$ coefficients with 
$\td{ij}=\td{ij}(k_2,q_1,q_2)$ are listed below.
\begin{eqnarray}
\mathbf{Box}_1^{(2)}
&=&-\epsbqa [ 8 \td{27}-8 \td{312}+24 \td{313} -4 (\td{23}-\td{26}+\td{33} -\td{39}) \qbs+\td{11} s - \td{12} s \nonumber \\
&+& 5 \td{13} s+8 \td{25} s-4 \td{26} s +4 \td{37} s-4 \td{39} s-\td{11} u - 7 \td{12} u+7 \td{13} u+4 \td{23} u \nonumber \\
&-& 8 \td{24} u+4 \td{25} u-4 \td{310} u + 4 \td{37} u ] -\epsbqb [8 \td{27}+16 \td{313} -4 (\td{23}-\td{26}+\td{33} \nonumber \\
&-&\td{39}) \qbs+4 \td{13} s + 4 \td{23} s + 4 \td{25} s-4 \td{26} s+4 \td{37} s -4 \td{39} s-\td{11} u-3 \td{12} u \nonumber \\
&+& 3 \td{13} u+8 \td{23} u-4 \td{24} u -4 \td{26} u-4 \td{310} u+4 \td{37} u ] - \epsbkb [ -8 \td{311}+24 \td{313} \nonumber \\
&-& (\td{11} -\td{12}+\td{13}+4 \td{25} -4 \td{26}+4 \td{33} - 4 \td{39}) \qbs +\td{11} s-\td{12} s+5 \td{13} s \nonumber \\
&+& 8 \td{25} s-4 \td{26} s+4 \td{37} s -4 \td{39} s-4 \td{12} u+4 \td{13} u + 4 \td{23} u-4 \td{24} u+4 \td{25} u \nonumber \\
&-&4 \td{26} u - 4 \td{310} u+4 \td{37} u ]
\end{eqnarray}
\begin{eqnarray}
\mathbf{Box}_2^{(2)}
&=&-\epsaqb [ -8 \td{27}-8 \td{313}+(\td{11} +3 \td{12}-3 \td{13}+4 \td{24} -4 \td{26}) u] \nonumber \\
&-& \epsakb [ 16 \td{311}-24 \td{312} + (\td{11}+3 \td{12}-3 \td{13} -4 \td{23}+4 \td{24}+4 \td{310} \nonumber \\
&-&  4 \td{37}-4 \td{38}+4 \td{39}) \qbs + 5 \td{11} s-5 \td{12} s+\td{13} s + 4 \td{21} s -4 \td{24} s+4 \td{25} s \nonumber \\
&-& 4 \td{26} s-8 \td{310} s+4 \td{35} s +4 \td{38} s-4 \td{12} u + 4 \td{13} u +4 \td{22} u-8 \td{24} u+8 \td{25} u \nonumber \\
&-& 4 \td{26} u-4 \td{310} u-4 \td{34} u + 4 \td{35} u+4 \td{36} u ]
\end{eqnarray}
\begin{eqnarray}
  \mathbf{Box}_b^{(2)} 
&=&-4 \td{27}-12 \td{312}+12 \td{313} -2 \td{33} \qbs-2 \td{38} \qbs+4 \td{39} \qbs -2 \td{0} s - \td{11} s-\td{12} s \nonumber \\
&+&\td{13} s+2 \td{25} s-2 \td{26} s -2 \td{310} s+2 \td{37} s+2 \td{38} s - 2 \td{39} s+2 \td{22} u+2 \td{23} u \nonumber \\
&-&4 \td{26} u -4 \td{310} u+2 \td{36} u + 2 \td{37} u
\end{eqnarray}
\begin{eqnarray}
\mathbf{Box}_q^{(2)}
&=&-8 \td{312} \epsaepsb+8 \td{313} \epsaepsb-8 \td{12} \epsakb \epsbkb + 8 \td{13} \epsakb \epsbkb \nonumber \\
&-& 12 \td{24} \epsakb \epsbkb+12 \td{25} \epsakb \epsbkb - 4 \td{34} \epsakb \epsbkb+4 \td{35} \epsakb \epsbkb \nonumber \\
&+& \td{11} \epsaqb \epsbkb \nonumber - \td{12} \epsaqb \epsbkb+\td{13} \epsaqb \epsbkb+4 \td{23} \epsaqb \epsbkb \nonumber \\
&+& 4 \td{25} \epsaqb \epsbkb-8 \td{26} \epsaqb \epsbkb-4 \td{310} \epsaqb \epsbkb + 4 \td{37} \epsaqb \epsbkb \nonumber \\
&-& \td{11} \epsakb \epsbqa-7 \td{12} \epsakb \epsbqa + 7 \td{13} \epsakb \epsbqa-4 \td{22} \epsakb \epsbqa \nonumber \\
&-& 8 \td{24} \epsakb \epsbqa + 4 \td{25} \epsakb \epsbqa+8 \td{26} \epsakb \epsbqa+4 \td{310} \epsakb \epsbqa \nonumber \\
&-& 4 \td{36} \epsakb \epsbqa+4 \td{23} \epsaqb \epsbqa-4 \td{26} \epsaqb \epsbqa - 4 \td{38} \epsaqb \epsbqa \nonumber \\
&+& 4 \td{39} \epsaqb \epsbqa-\td{11} \epsakb \epsbqb - 3 \td{12} \epsakb \epsbqb+3 \td{13} \epsakb \epsbqb \nonumber \\
&+& 8 \td{23} \epsakb \epsbqb - 4 \td{24} \epsakb \epsbqb-4 \td{26} \epsakb \epsbqb-4 \td{310} \epsakb \epsbqb \nonumber \\
&+& 4 \td{37} \epsakb \epsbqb+4 \td{23} \epsaqb \epsbqb-4 \td{26} \epsaqb \epsbqb + 4 \td{33} \epsaqb \epsbqb \nonumber \\
&-& 4 \td{39} \epsaqb \epsbqb-2 \td{33} \epsaepsb \qbs - 2 \td{38} \epsaepsb \qbs+4 \td{39} \epsaepsb \qbs \nonumber \\
&+& \frac{1}{2}\td{11} \epsaepsb s - \frac{1}{2}\td{12} \epsaepsb s+\frac{1}{2}\td{13} \epsaepsb s+2 \td{25} \epsaepsb s \nonumber \\
&-& 2 \td{26} \epsaepsb s-2 \td{310} \epsaepsb s+2 \td{37} \epsaepsb s +2 \td{38} \epsaepsb s \nonumber \\
&-& 2 \td{39} \epsaepsb s+2 \td{22} \epsaepsb u +2 \td{23} \epsaepsb u-4 \td{26} \epsaepsb u \nonumber \\
&-& 4 \td{310} \epsaepsb u + 2 \td{36} \epsaepsb u+2 \td{37} \epsaepsb u
\end{eqnarray}
%

In the above expressions, the finite $\td{ij}$ functions are obtained 
by standard Passarino-Veltman recursion relations~\cite{Passarino:1978jh} 
from the finite parts of the basic scalar integrals. 
For the virtual corrections considered, only the one-mass 
box~\cite{Bern:1993kr,Papadopoulos:1981ju}, is needed.  
Specifically, we need the case in which $k_{1}^2 = k_{2}^2 = q_{1}^2=0$ 
and $ q_{2}^{2} \ne 0$. Here $k_{i}$ and $q_{i}$
with $i=1,2$ are the external four momenta. The one-mass box in the 
unphysical region, $-s>0, -t >0, -q_{2}^2 > 0 $ is,
\beqn
D_{0}(k_2,q_2,q_1) &=& \int \frac{d^{d} k}{i \pi^2} 
\frac{1}{[k^2 ] [ (k-k_2)^2 ] [ (k-k_2 - q_2)^2 ] [(k-k_2 - q_2 - q_1)^2 ]} \\ \nonumber 
 &=& \pi^{-\epsilon} (\mu^2)^{-\epsilon} \Gamma(1+\epsilon) \\ \nonumber 
& \cdot &
\left \{ \frac{2}{st} \frac{1}{\epsilon^2} + \frac{2}{st} \frac{1}{\epsilon}
\left [ \ln \left ( \frac{-q_{2}^{2}}{\mu^2} \right ) - \ln \left ( \frac{-s}{\mu^2} \right ) - 
\ln \left (\frac{-t}{\mu^2} \right) \right]
+ \widetilde{D}_{0}(k_2,q_2,q_1) + \mathcal{O}(\epsilon) \right \},
\eeqn
where, 
\beqn
\widetilde{D}_{0}(k_2,q_2,q_1) &=& \frac{1}{st} 
\left [ \ln^2 \left ( \frac{-s}{\mu^2} \right)  + \ln^2 \left ( \frac{-t}{\mu^2} \right ) 
-\ln^2 \left ( \frac{-q_{2}^{2}}{\mu^2} \right ) \right. \\ \nonumber 
&-& \left .  \ln \left ( \frac{-s}{\mu^2} \right) + \ln \left( \frac{-t}{\mu^2} \right ) 
- 2 ~{\rm Li}_{2} \left ( 1 - \frac{q_{2}^{2}}{t} \right ) 
-2 ~{\rm Li}_{2} \left ( 1 - \frac{q_{2}^{2}}{s} \right ) - \frac{2 \pi^2}{3} \right ].
\eeqn
The Mandelstam variables, $s$ and $t$, are defined in Eq.~(\ref{eq:mns}).

For the present application, the invariant, $q_{2}^{2}$ is always 
space-like while the Mandelstam 
invariants, $s$ and $t$, may either be time-like or space-like. 
Results for physical kinematic regions can
be obtained by analytic continuation by replacing the time-like invariant 
by $t \rightarrow t + i 0^{+}$ or $s \rightarrow s + i 0^{+}$.

In addition, to the one-mass box, we also require expressions for the 3-point 
and 2-point scalar
integrals in $d=4-2 \epsilon$ space-time dimensions.
For the 3-point scalar integral,
\beqn
C_{0}(p_{1}^2,p_{2}^2,(p_{1} + p_{2})^2) &=&   \int \frac{d^d k}{i \pi^2} 
  \frac{1}{[-k^2-i0^{+}][-(k+p_1)^2 -i0^{+}]} \nonumber \\
&\times& \frac{1}{[-(k+p_1+p_2)^2-i0^{+}]},
\eeqn
two cases are needed. Here $p_{1}$ and $p_{2}$ represent the external 
outward flowing four momenta.
\begin{itemize}
\item[(a).] For the two--mass triangle, either, $p_{1}^{2}=0$ or $p_{2}^{2}=0$ 
and $p_{3}^2=(p_{1} + p_{2})^2 \ne 0$.
\beqn
C_{0}(p_{1}^2,0,p_{3}^2)&=& \pi^{-\epsilon} (\mu^2)^{-\epsilon} \Gamma(1+\epsilon)\\ \nonumber 
&\cdot& \left \{ \frac{1}{-p_{3}^2-p_{1}^2} \left(\ln \left(\frac{-p_{3}^{2}-i0^{+}}{\mu^2} \right)-\ln \left(\frac{-p_{1}^2-i0^{+}}{\mu^2} \right) \right) \frac{1}{\epsilon} \right. \nonumber \\ 
&+& \left. \widetilde{C}_{0}(p_{1}^2,p_{3}^2) + \mathcal{O}(\epsilon)\right \} \nonumber \\
\widetilde{C}_{0}(p_{1}^2,p_{3}^2)&=& \frac{1}{2} \frac{1}{-p_{3}^2-p_{1}^2} 
\left( \ln^2 \left(\frac{-p_{1}^2-i0^{+}}{\mu^2} \right) - \ln^2 \left(\frac{-p_{3}^2-i0^{+}}{\mu^2} \right) \right) 
\eeqn
\item[(b).] For the one--mass triangle, $p_{1}^2 = p_{2}^2 =0$ and
$p_{3}^2=(p_{1} + p_{2})^2 \ne  0$. 
\beqn
C_{0}(0,0,p_{3}^2)&=& \pi^{-\epsilon} (\mu^2)^{-\epsilon} \Gamma(1+\epsilon)
\left \{ \frac{1}{-p_{3}^2} \frac{1}{\epsilon^2}\right. \\ \nonumber 
&-& \frac{1}{-p_{3}^2} \ln \left(\frac{-p_{3}^2-i0^{+}}{\mu^2} \right) \frac{1}{\epsilon}
+ \left. \widetilde{C}_{0}(p_{3}^2) + \mathcal{O}(\epsilon)\right \}\\
\widetilde{C}_{0}(p_{3}^2)&=& -\frac{\pi^2}{6}\frac{1}{-p_{3}^2}
+ \frac{1}{-p_{3}^2}\frac{1}{2} \ln^2 \left(\frac{-p_{3}^2-i0^{+}}{\mu^2} \right) 
\eeqn

\end{itemize}
The scalar 2-point integral is
\beqn
B_{0}(q^2) &=&\int \frac{d^{d} k}{ i \pi^2} \frac{1}{[-k^2- i0^{+}] [-(k-q)^2 - i0^{+} ]} \\ \nonumber 
                    &=& \pi^{-\epsilon} (\mu^2)^{-\epsilon} \Gamma(1+\epsilon)
                    \left [ \frac{1}{\epsilon} + \widetilde{B}_{0}(q^2) + \mathcal{O}(\epsilon) \right ]
\eeqn
with
\beqn
\widetilde{B}_{0}(q^2) = 2 - \ln \frac{-q^2 - i 0^{+}}{\mu^2} \, .
\eeqn

\section{Cross section formulas}\label{app:B}
In this appendix we give cross section formulas for processes of the type, 
\beq
g(p_a) + Q(p_b) \rightarrow q(p_1) + \bar{q}(p_3) + Q(p_2) + H(P).
\eeq
Results for the crossed process 
$q(p_a) + Q(p_b) \rightarrow q(p_1)+Q(p_2)+g(p_3)  + H(P)$ were already 
given in Section~\ref{sec:calc}.
The finite three parton NLO cross section that results from the cancellation 
of the $1/\eps^2$ and $1/\eps$ poles of the virtual corrections with those 
of the insertion operator, $\mathbf{I}(\eps)$, is
\beqn
\sigma^{NLO}_{3}(g Q \rightarrow q \bar{q} Q H) &=& 
\int_{0}^{1} dx_{a} \int_{0}^{1} dx_{b} f_{g/p}(x_a,\mu_F) 
f_{Q/p}(x_b,\mu_F) \\ \nonumber 
&\times &\frac{1}{2 \hat{s}} d \Phi_{4}(\pa,\pb) 
F_{J}^{(3)}(\p1,\p2,\p3,P;\pa,\pb) \\ \nonumber 
&\cdot &\overline{\sum_{\rm colors}}\;
\left \{ |\mc{M}_{3}(1_{q},2_{Q},3_{\bar{q}},a_{g},b_{Q})|^2 
\left ( 1 + \frac{\alpha_s(\mu^2)}{2 \pi} 
\left(K_{\rm born} + F(s_{13},s_{a3},s_{a1})\right) \right) \right. \\ \nonumber 
&+ & \left. 2~\Re[ 
\widetilde{\mc{M}}_{3}^{virt}(1_{q},2_{Q},3_{\bar{q}},a_{g},b_{Q}) 
\mc{M}_{3}^{*}(1_{q},2_{Q},3_{\bar{q}},a_{g},b_{Q})]\right \} \, ,
\eeqn
with 
\beqn
\widetilde{\mc{M}}_{3}^{virt}(1_{q},2_{Q},3_{\bar{q}},a_{g},b_{Q})=
t^{c_{a}} \delta_{i_{1} i_{3}} 
\widetilde{\mc{M}}_{V}(\p{1},-\p{a},p_{a13};\eps_{a},
h(\p{b} \tau_{b},\p{2} \tau_{2})) \, .
\eeqn
The Born level matrix element squared is
\beqn
\overline{\sum_{\rm colors}}\;
|\mc{M}_{3}(1_{q},2_{Q},3_{\bar{q}},a_{g},b_{Q})|^2=\frac{C_{F} N}{N^2-1} 
|\mc{A}_{3}(1_{q},a_{g},3_{\bar{q}};2_{Q},b_{Q})|^2 \,
\eeqn
with 
\beqn
\mc{A}_{3}(1_{q},a_{g},3_{\bar{q}};2_{Q},b_{Q}) = 
\mc{M}_{B}(\p{1},-\p{a},p_{a13};\eps_{a},h(\p{b} \tau_{b},\p{2} \tau_{2}))\, .
\eeqn
The finite collinear contribution is
\beqn
\sigma_{3,{\rm col}}^{NLO}(gQ \rightarrow q \bar{q} Q H)& =& 
\int_{0}^{1} dx_{a} \int_{0}^{1} dx_{b} 
\frac{1}{2 \hat{s}} d \Phi_{4}(\pa , \pb)F_{J}^{(3)}(\p1,\p2,\p3;\pa,\pb) 
\\ \nonumber 
&\cdot & \left \{  f_{g/p}(x_{a};\mu_F) f_{Q/p}^{2,b}(x_{b};\mu_F,\mu_R) 
\right. \\ \nonumber 
&+&\left. \frac{1}{2}\left( f_{g/p}^{1,a}(x_{a};\mu_F,\mu_R) + 
f_{g/p}^{3,a}(x_{a};\mu_F,\mu_R)\right ) f_{Q/p}(x_{b};\mu_F)\right \} 
\\ \nonumber
&\cdot& \frac{C_F N}{N^2-1}
|\mc{A}_{3}(1_{q},a_{g},3_{\bar q};2_{Q},b_{Q})|^2 
\eeqn
with
\begin{eqnarray}
f^{i,a}_{g/p}(x_a;\mu_F,\mu_R)
&=&\frac{\alpha_s(\mu_R)}{2 \pi} \int_{x_a}^{1} \frac{dz}{z}
\left \{\sum_{q} \left[ f_{q/p}\left (\frac{x_a}{z};\mu_F \right) 
+f_{\bar{q}/p}\left (\frac{x_a}{z};\mu_F \right) \right]A^{i,a}_{qg}(z)\right.     \nonumber \\   
&+&\left [f_{g/p}\left (\frac{x_a}{z};\mu_F \right) - z f_{g/p}(x_a;\mu_F) \right] B^{i,a}_{gg}(z) \\ \nonumber 
&+&\left. f_{g/p}\left (\frac{x_a}{z};\mu_F \right) C^{i,a}_{gg}(z) \right \} + \frac{\alpha_s(\mu_R)}{2 \pi} f_{g/p}(x_a;\mu_F) D^{i,a}_{gg}(x_a) \nonumber ,
\end{eqnarray}
with kernels,
\begin{eqnarray}
A^{i,a}_{qg}(z)& =&C_F \left[\frac{1+(1-z)^2}{z}\ln \frac{2p_a p_i(1-z)}{\mu_F^2 z}+z \right] \\
B^{i,a}_{gg}(z)&=&C_A\left [\frac{2}{1-z} \ln \frac{2p_a p_i(1-z)}{\mu_F^2}-
\frac{3}{2}\frac{1}{1-z}\right ], \\ 
C^{i,a}_{gg}(z)&=& 2 C_A \left [\left(\frac{1-z}{z}-1+z(1-z)\right)\ln \frac{2p_a p_i(1-z)}{\mu_F^2 z} -
\frac{1}{1-z} \ln z \right] ,\\
D^{i,a}_{gg}(x)&=&2 C_A \ln(1-x) \ln \frac{2p_a p_i}{\mu_F^2}+
\gamma_g  \ln \frac{2p_a p_i}{\mu_F^2} \nonumber \\
&+&C_A \left (\frac{2 \pi^2}{3}-\frac{50}{9}+\ln^2(1-x) \right)\\
&+&\frac{16}{9}T_R N_f
-\frac{3}{2}C_A-\frac{3}{2}C_A\ln(1-x) \nonumber \, .
\end{eqnarray}